\documentclass[authoryear,a4paper,review]{elsarticle} 
\pdfoutput=1

\journal{Journal of Futures Markets}

\usepackage{amsmath,amssymb,natbib,graphicx,tikz,hyperref,thmtools,soul,ifthen,enumerate,setspace,array}
\usepackage[left=1in,right=1in,top=1.4in,bottom=1.2in]{geometry}

\hypersetup{
	colorlinks=true,
	linkcolor=red,
	citecolor=blue,
	urlcolor=blue,
	bookmarksnumbered=true,
	pdfstartview=
}

\newcommand{\qtext}[2][\quad]{#1\text{#2}#1}

\newcommand{\mat}[1]{\boldsymbol{#1}}
\newcommand{\vect}[1]{\boldsymbol{#1}}
\newcommand{\xtopxinv}{(\mat{X}^\top \mat{X})^{-1}}

\newcommand{\lsf}[2]{\ifthenelse{\equal{#2}{}}{#1}{#1^{[#2]}}}
\newcommand{\lsv}[2]{\ifthenelse{\equal{#2}{}}{\boldsymbol{#1}}{\boldsymbol{#1}^{[#2]}} }
\newcommand{\lsvel}[3]{\ifthenelse{\equal{#2}{}}{#1_{#3}}{#1^{[#2]}_{#3}} }

\newcommand{\Cov}{\mathrm{Cov}}
\newcommand{\LSM}{\textsc{lsm}}

\newcommand{\LOOLSM}{\textsc{loo}}
\newcommand{\swap}[1]{\textsc{Sw}^{#1\text{y}}}
\newcommand{\ExpMC}{\mathbb{E}_\omega}
\newcommand{\ProbMC}{\mathbb{P}_\omega}
\newcommand{\ExpN}{\mathbb{E}_n}

\newcolumntype{L}[1]{>{\raggedright\let\newline\\\arraybackslash\hspace{0pt}}m{#1}}
\newcolumntype{C}[1]{>{\centering\let\newline\\\arraybackslash\hspace{0pt}}m{#1}}
\newcolumntype{R}[1]{>{\raggedleft\let\newline\\\arraybackslash\hspace{0pt}}m{#1}@{\hspace{1em}}}

\newtheorem{assumption}{Assumption}
\newtheorem{lemma}{Lemma}
\newproof{proof}{Proof}

\begin{document}

\begin{frontmatter}

\title{Leave-one-out least squares Monte Carlo algorithm\\ for pricing Bermudan options}

\author[jwoo]{Jeechul Woo}
\ead{jwoo@modulitech.com}
\address[jwoo]{Moduli Technologies, Springfield, Illinois, USA}

\author[stan]{Chenru Liu}
\ead{liucr@stanford.edu}
\address[stan]{Department of Management Science and Engineering, Stanford University, Stanford, California, USA}

\author[phbs]{Jaehyuk Choi\corref{corrauthor}}
\ead{jaehyuk@phbs.pku.edu.cn}
\address[phbs]{Peking University HSBC Business School, Shenzhen, China}

\cortext[corrauthor]{Corresponding author \textit{Tel:} +86-755-2603-0568, \textit{Address:} Peking University HSBC Business School, University Town, Nanshan District, Shenzhen 518055, China}

\begin{abstract}
The least squares Monte Carlo (LSM) algorithm proposed by \citet{longstaff2001val} is widely used for pricing Bermudan options. The LSM estimator contains undesirable look-ahead bias, and the conventional technique of avoiding it requires additional simulation paths. We present the leave-one-out LSM (LOOLSM) algorithm to eliminate look-ahead bias without doubling simulations. We also show that look-ahead bias is asymptotically proportional to the regressors-to-paths ratio. Our findings are demonstrated with several option examples in which the LSM algorithm overvalues the options. The LOOLSM method can be extended to other regression-based algorithms that improve the LSM method.
\end{abstract}

\date{\today}

\begin{keyword}
	American option, Bermudan option, Least squares Monte Carlo, Look-ahead bias, Leave-one-out-cross-validation
\end{keyword}
\end{frontmatter}


\section{Introduction} \label{sec:intro} \noindent
Derivatives with early exercise features, such as American and Bermudan options, are popular in the market. Nonetheless, pricing these options is a difficult problem in the absence of closed-form solutions, even in the simplest case of American options on a single asset. Researchers have thus developed various numerical methods for pricing that largely fall into two categories: the lattice-based and simulation-based approaches.

In the lattice-based approach, pricing is performed on a dense lattice in the state space by valuing the options at each point of the lattice using suitable boundary conditions and the mathematical relations among neighboring points. Examples include the finite difference scheme~\citep{brennan1977valuation}, binomial tree~\citep{cox1979option}, and its multidimensional generalizations~\citep{boyle1988lattice,boyle1989num,he1990lattice}. These methods are known to work well in low-dimensional problems. However, they become impractical in higher-dimensional settings, mainly because the lattice size grows exponentially as the number of state variables increases. This phenomenon is commonly referred to as the curse of dimensionality. 

In the simulation-based approach, the price is calculated as the average of the option values over simulated paths, each representing a future realization of the state variables with respect to the risk-neutral measure. While the methods in this category are not challenged by dimensionality, they entail finding the optimal exercise rules. Several simulation-based methods propose various approaches for estimating the continuation values as conditional expectations. Equipped with stopping time rules, they calculate the option price by solving a dynamic programming problem whose Bellman equation is essentially the comparison between the continuation values and exercise values. In comparison to the randomized tree method~\citep{broadie1997pricing} and stochastic mesh method~\citep{broadie2004mesh}, regression-based methods~\citep{carriere1996val,tsitsiklis2001reg,longstaff2001val} are  computationally tractable. Because the methods use regression to estimate the continuation values from the simulated paths, their computation time is linear not only in the number of simulated paths, but also in the number of exercise times. Among the regression-based methods, the least squares Monte Carlo (LSM) algorithm proposed by \citet{longstaff2001val} is the most popular for its simplicity and efficiency. \citet{fu2001pricing} and \citet{glasserman2003mc_ch8} provide comprehensive reviews of the implementation and comparison of simulation-based methods.

The LSM method is essential for pricing callable structured notes whose coupons have complicated dependency on other underlying assets such as equity prices, foreign exchange rates, and benchmark interest swap rates. Because a multi-factor model is required for the underlying assets and the yield curve term structure, using Monte Carlo simulation along with the LSM method is inevitable for pricing and risk-managing such notes. Like many previous studies on this topic~\citep{kolodko2006iter,beveridge2013ppi}, this study is motivated and developed in the context of callable structured notes.

\subsection{Biases in the LSM method} \noindent
In simulation-based methods including the LSM method, there are two main sources of bias, which run in opposite directions. Low-side bias is related to suboptimal exercise decisions owing to various approximations adopted in the method. In the LSM method, for example, finite basis functions cannot fully represent the conditional payoff function. The resulting exercise policy deviates from the most optimal one, leading to a lower option price. For this reason, it is also called \textit{suboptimal} bias. 
High-side bias comes from using one simulation set for both the exercise decision and the payoff valuation. As explained by \citet{broadie1997pricing}, this practice creates a fictitious positive correlation between exercise decisions and future payoffs; the algorithm is more likely to continue (exercise) precisely when the \textit{future} payoff in the simulation is higher (lower). For this reason, it is called \textit{look-ahead} or \textit{foresight} bias. The LSM estimator has both low- and high-side biases; hence, \citet{glasserman2003mc_ch8} calls it an interleaving estimator. Other simulation estimators in the literature are typically either low-biased or high-biased. For example, \citet{broadie1997pricing} carefully construct both low- and high-biased estimators to form a confidence interval for the true option price. 

In callable structured note markets, look-ahead bias is more dangerous than suboptimal bias, and look-ahead bias mixed with suboptimal bias is a significant drawback of the LSM estimator. This is closely related to the typical roles of the market players. The financial institutions that issue notes are the effective buyers of the Bermudan option to redeem notes early. On the contrary, investors, be they individual or institutional, are the effective sellers of the option. Investors receive the option premium in the form of an enhanced note yield compared to noncallable notes with the same structure.\footnote{The buyers and sellers of the Bermudan option rarely change in the market. Financial institutions hardly issue \textit{puttable} notes, where they sell the option to investors. Puttable notes are not attractive to investors because the yield would be lower (i.e., investors pay the option premium).} As a result, option buyers (financial institutions) act as pricing agents. Because buyers have to risk-manage and optimally exercise the option, they typically use the LSM method. Option sellers (investors) usually hold the note until maturity without hedging and, therefore, are less sensitive to accurate valuation. From option buyers' perspective, look-ahead bias is malicious because it wrongly inflates the option premium they pay. No matter how well the option is delta-hedged, the option value attributed from look-ahead bias shrinks to zero when the position is near the maturity or the early exercise because there is no more \textit{future} to look into by then. Suboptimal bias, on the contrary, is benign. Although it deflates the option value, the gain realized through delta-hedging under the suboptimal exercise policy is just as much as the deflated option value. In short, option buyers \textit{get what they pay for}. The only downside of suboptimal bias for buyers is to lose trades to competitors who bid a higher (more optimal) option premium. Therefore, buyers prefer the low-biased estimator to ensure that the premium they pay is lower than the true value. However, look-ahead bias mixed in the LSM method makes a \textit{conservative} valuation difficult for buyers.

A standard technique for eliminating look-ahead bias is calculating the exercise decision using an additional independent set of Monte Carlo paths, thereby eliminating the correlation between the exercise decision and simulated payoff. While this two-pass approach removes look-ahead bias, it comes at the cost of doubling the computational cost, which is already heavy because the simulation of stochastic processes frequently requires the time-discretized Euler scheme. The design of the LSM estimator to include the biases in both directions primarily aims to retain computational efficiency rather than raise accuracy by letting these two biases partially offset. Moreover, \citet{longstaff2001val} claim that the look-ahead bias of the LSM estimator is negligible by presenting a single-asset put option case tested with the two-pass simulation as supporting evidence. In this regard, the LSM estimator has been considered to be low-biased.

However, researchers and practitioners have raised concerns that look-ahead bias may not be small in multiasset problems, where the simulation must be the last resort method. The numerical results of \citet{letourneau2014refining} and \citet{fabozzi2017hetero} show that look-ahead bias increases when the simulation size becomes smaller or the polynomial order of the basis becomes higher. \citet{carriere1996val} and \citet{fries2005foresight} remark the same.
Practitioners in the structured notes market also observe that when higher-order regression variables are used to capture the exercise boundary better (i.e., reduce suboptimal bias), look-ahead bias also increases. Given the desire to keep the one-pass LSM implementation for computational efficiency, they are reluctant to include higher-order terms in the LSM regression for fear of overpricing. It is possible to check the validity of the LSM price against several methods to estimate both the lower and the upper bounds of American options based on policy iteration~\citep{kolodko2006iter,beveridge2013ppi} and duality representation~\citep{haugh2004pricing,andersen2004primal}, respectively. However, their computational cost is too heavy to be used in day-to-day pricing and risk management, as nested simulations are required. Therefore, it is of significant practical importance to understand the magnitude of look-ahead bias in the LSM estimator and develop an efficient algorithm for removing it. 

\subsection{Contribution of this study} \noindent
This study presents an efficient approach for removing look-ahead bias, motivated by the cross-validation practice in statistical learning. Standard practice is to separate the data sets for training and testing to avoid overfitting. In this context of statistical learning, look-ahead bias is overfitting caused by using the same data set for both training (i.e., the estimation of the exercise policy) and testing (i.e., the valuation of the options). Similarly, using an independent simulation set for the exercise policy corresponds to the hold-out method, one of the simplest cross-validation techniques.

Among advanced cross-validation techniques, we recognize that leave-one-out cross-validation (LOOCV) fits with the LSM method. When making a prediction for a sample, LOOCV trains the model with all samples except one, thereby separating the data set for testing in the most minimal way. In linear regression, it is well known that the corrections from the full regression on all samples can be computed altogether with a simple linear algebra operation~\citep[\S~7.10]{hastie2009ESL}. Therefore, our new leave-one-out LSM (LOOLSM) algorithm eliminates look-ahead bias from the LSM method without incurring an extra computational cost. The LOOLSM method can thus be understood as an extension of the low-biased estimator of \citet{broadie1997pricing} in that self-exclusion is conducted on all simulation paths rather than on each state separately. By using the LOOLSM method, practitioners can reliably obtain the low-biased price---even with higher-order regression basis functions. The LOOLSM algorithm can also be applied along with other regression methods proposed to improve least squares regression~\citep{tompaidis2014altols,chen2019var, ibanez2018opt,fabozzi2017hetero,belomestny2011pricing,ludkovski2018kriging}.

Furthermore, this study contributes to the line of research dealing with the convergence of the LSM algorithm, which is a problem of fundamental importance given the popularity of the method. Several authors theoretically analyze the convergence of the LSM method; \citet{clement2002anal} prove the convergence of the LSM price for a fixed set of regressors based on the central limit theorem. \citet{stentoft2004conv} analyzes the convergence rate of the continuation value function when the number of regressors $M$ also goes to infinity. \citet{glasserman2004number} discuss how quickly the simulation size $N$ has to grow relative to $M$ to achieve uniform convergence. \citet{zanger2018conv} estimates the stochastic component of the error for a general class of approximation architecture.

Previous studies focus primarily on the convergence of the continuation value functions in the $L^2$ space. By construction, they do not analyze the convergence rate of look-ahead bias specific to the LSM method, in which the estimated continuation value functions are evaluated for the training samples. We bridge this research gap. In particular, we formulate look-ahead bias as the difference between the LSM and LOOLSM prices, with which we theoretically analyze its convergence rate and derive the upper bounds in Theorem~\ref{theorem:3.1}. Empirically, the formulation provides a way to measure look-ahead bias more robust to Monte Carlo noise. We conduct numerical studies for options whose true prices are known and obtain results consistent with our theoretical findings.

To the authors' best knowledge, previous works estimating look-ahead bias, theoretically or empirically, are scarce, and those correcting such bias are rare. \citet{carriere1996val} predicts that the high-side bias of the estimator asymptotically scales to $1/N + O(1/N^2)$. Our analysis and simulation results not only reaffirm this observation, but also show that any realistic look-ahead bias decays at the regressors-to-paths ratio $M/N$ at least. \citet{fries2005foresight,fries2008foresight} formulates look-ahead bias as the price of the option on the Monte Carlo error and derives the analytic correction terms from the Gaussian error assumption. Compared with these studies, our LOOLSM method does not depend on any model assumption and more accurately targets look-ahead bias in the LSM setting.

Beyond the American option pricing, our new method can be applied to various stochastic control problems in finance where least squares regression approximates the optimal strategy. For examples, see \citet{huang2016reg} for variable annuities, \citet{nadarajah2017comp} for energy real options, and \citet{bacinello2010reg} for life insurance contracts.

The rest of the paper is organized as follows. In \S~\ref{sec:loolsm}, we describe the LSM pricing framework and introduce the LOOLSM algorithm. In \S~\ref{sec:labias}, we analyze the convergence rate of look-ahead bias in the LSM method. In \S~\ref{sec:num}, the numerical results are presented, and, in \S~\ref{sec:ext}, the LOOLSM method is extended to other regression estimators. Finally, \S~\ref{sec:con} concludes.

\section{Method} \label{sec:loolsm} \noindent
This section briefly reviews American option pricing, primarily to develop our method later. For a detailed review, see \citet{glasserman2003mc_ch8}. We first introduce the conventions and notations used in the rest of the paper:
\begin{itemize}
	\item The option can be exercised at a discrete time set $\{0 < t_1 < \cdots < t_I=T\}$. It is customary to assume that $t_0=0$ is \textit{not} an exercise time because the buyer would not buy an option if it is optimal to exercise it immediately.
	\item $S(t) = (S_1(t), \cdots, S_J(t))$ denotes the Markovian state vector at time $t$. We denote the value at $t_i$ by $\lsf{S}{i} = S(t_i)$. The state vector $\lsf{S}{i}$ usually includes the underlying asset prices, but it is not limited to them. In Section~\ref{ssec:cmsra}, we present a numerical example (Case~4) where $\lsf{S}{i}$ includes stochastic volatility in addition to forward interest rates.
	\item $\lsf{Z}{i}(s)$ denotes the expected payout given the option is exercised at time $t_i$ and state $\lsf{S}{i}=s$. It is discounted to the present time $t_0=0$. For the classical single-stock put option with strike price $K$ and constant risk-free rate $r$ (see Case~1 in Section~\ref{ssec:put}), it is defined as $\lsf{Z}{i}(s) = e^{-r\,t_i}\max(K - s_1,0)$. However, the LOOLSM algorithm is still valid in the models with a stochastic risk-free rate or interest rate term-structure model (see Case~4 in Section~\ref{ssec:cmsra}). In general, the exact payout of Bermudan options may depend on the path of $S(t)$ \textit{after} exercise time $t_i$; hence, the expected payout.
	\item $\lsf{V}{i}(s)$ and $\lsf{C}{i}(s)$ denote the discounted option values at time $t_i$ and state $\lsf{S}{i}=s$ given that the option was not exercised up to (and at) $t_{i-1}$ and $t_i$, respectively. Prior studies commonly refer to $\lsf{C}{i}(s)$ as the \textit{continuation value}.
\end{itemize}
The exercise time index $[i]$ or the time dependency $(t)$ may be omitted when it is clear from the context.

We can formulate the valuation of options with early exercise features as a maximization problem of the expected future payoffs over all possible choices of discrete stopping times taking values in $\{1,\cdots,I\}$:
\begin{equation} \label{eq:stoppingtime}
\lsf{V}{0}(s) = \max_{\tau \in \mathcal{T}}~\mathbb{E}[ \lsf{Z}{\tau}(\lsf{S}{\tau}) \,|\, \lsf{S}{0}=s ].
\end{equation}
This is equivalent to a dynamic programming problem using the continuation value. Since $\lsf{C}{i}(s)$ and $\lsf{V}{i+1}(s)$ are related by
$$
\lsf{C}{i}(s) = \mathbb{E} [\,\lsf{V}{i+1}(\lsf{S}{i+1}) \, | \, \lsf{S}{i}=s\,]
\qtext{for} 0 \le i < I,
$$ 
we calculate the discounted option value at $t_i$ by backward induction,
\begin{equation} \label{eq:bellman}
\lsf{V}{i}(s) = \max(\lsf{C}{i}(s),\, \lsf{Z}{i}(s)). 
\end{equation}
This effectively means that the option continues at $t_i$ if $\lsf{C}{i}(s) \ge \lsf{Z}{i}(s)$ and is exercised otherwise. For consistency, we assume $\lsf{Z}{0}(s) = \lsf{C}{I}(s) = -\infty$ to ensure that $\lsf{V}{0}(s) = \lsf{C}{0}(s)$ (i.e., must continue at $t_0=0$) and $\lsf{V}{I}(s) = \lsf{Z}{I}(s)$ (i.e., must exercise at $t_I=T$ if not before). 
Therefore, we express the optimal stopping time $\tau$ in terms of $\lsf{C}{i}(s)$ and $\lsf{Z}{i}(s)$ as
$$\tau = \inf \{0< i\le I: \lsf{C}{i}(\lsf{S}{i}) < \lsf{Z}{i}(\lsf{S}{i}) \}.
$$

To see how the pricing works in the simulation setting, we further introduce the following conventions and notations:
\begin{itemize}
	\item We generate $N$ simulation paths of $\lsf{S}{i}$ ($1\le i\le I$) with the initial value $\lsf{S}{0}$. We denote the $n$-th simulation value of $\lsf{S}{i}$ by $\lsvel{S}{i}{n}$. 
	\item $\lsf{X}{i}(s) = (1, f_1(s), \cdots, f_{M-1}(s))$ denotes the set of $M$ basis functions at time $t_i$ and state $\lsvel{S}{i}{}=s$.
	\item The $N$-by-$M$ matrix $\lsv{X}{i}$ is the simulation result of $\lsf{X}{i}(s)$. The $n$-th row of $\lsv{X}{i}$, denoted by $\lsv{x}{i}_n$, corresponds to $\lsf{X}{i}(\lsvel{S}{i}{n})$. We assume the basis functions are \textit{diverse} enough to ensure that $\lsv{X}{i}$ has full column rank, $M$.
	\item The function $\lsf{\hat{C}}{i}(s)$ is an estimation of $\lsf{C}{i}(s)$ obtained from the simulation set, $\lsv{X}{i}$. 
	\item The length-$N$ column vectors, $\lsv{C}{i}$ and $\lsv{Z}{i}$, are the simulation values of $\lsf{\hat{C}}{i}(s)$ and $\lsf{Z}{i}(s)$, respectively. We denote the $n$-th elements by $\lsvel{C}{i}{n} = \lsf{\hat{C}}{i}(\lsvel{S}{i}{n})$ and $\lsvel{Z}{i}{n} = \lsf{Z}{i}(\lsvel{S}{i}{n})$.
	\item The vector $\lsv{Y}{i}$ is the length-$N$ column vector consisting of the option payout at the stopping time along the simulated paths, conditional on that the option was not exercised before $t_i$. We denote the $n$-th element by $\lsvel{Y}{i}{n}$ and it is equal to $\lsvel{Z}{\tau}{n}$ for some $i\le \tau\le I$. 
	\item For other variables to be defined later, we use the subscript $n$ and superscript $[i]$ consistently to denote the value of the $n$-th path at $t=t_i$.
	\item We use two types of expectation. In the first, we denote the expectation over the $N$ paths in one simulation set by $\ExpN$. In the second, we denote the expectation over repeated simulations by $\ExpMC$.
\end{itemize}

Following the stopping time formulation \eqref{eq:stoppingtime}, we compute $\lsv{Y}{i}$ as a path-wise backward induction step: $\lsvel{Y}{I}{n} = \lsvel{Z}{I}{n}$ and
\begin{equation} \label{eq:backind1}
\lsvel{Y}{i}{n} = \begin{cases}
\lsvel{Z}{i}{n} & \text{if $\lsvel{Z}{i}{n} > \lsvel{C}{i}{n}$}  \\
\lsvel{Y}{i+1}{n} & \text{if $\lsvel{Z}{i}{n} \le \lsvel{C}{i}{n}$} 
\end{cases} \;
= \mathcal{I}[\lsvel{C}{i}{n} \ge \lsvel{Z}{i}{n}]\cdot (\lsvel{Y}{i+1}{n} - \lsvel{Z}{i}{n}) + \lsvel{Z}{i}{n} 
\qtext{for} 0\le i < I,
\end{equation}
where $\mathcal{I}[\cdot]$ is the indicator function equal to 1 if the condition is satisfied and 0 otherwise. Many authors adopt this backward induction approach, notably \citet{tilley1993val}, \citet{carriere1996val}, and \citet{longstaff2001val}.\footnote{In an alternative backward induction formulation based on \eqref{eq:bellman},
	\begin{equation*} 
		\lsvel{Y'}{i}{n} = \max(\lsvel{C}{i}{n},\,\lsvel{Z}{i}{n}) = \mathcal{I}[\lsvel{C}{i}{n} \ge \lsvel{Z}{i}{n}] \cdot (\lsvel{C}{i}{n} - \lsvel{Z}{i}{n}) + \lsvel{Z}{i}{n},
	\end{equation*}
	which some authors such as \citet{carriere1996val} and \citet{tsitsiklis2001reg} adopt. However, we do not consider this approach. \citet{carriere1996val}, \citet{longstaff2001val}, and \citet{stentoft2014value} report that this alternative approach results in a bias significantly higher than the former approach, \eqref{eq:backind1}. See \citet{stentoft2014value} for the detailed comparison of the two approaches. \label{fn:alt}
}
In the final step of the backward induction, we calculate the option price estimate at $t_0=0$ as the average option value over the simulated paths:
\begin{equation} \label{eq:2.2}
\lsf{\hat{V}}{0} = \ExpN[\lsvel{Y}{0}{n}] = \frac1N \sum_{n=1}^N \lsvel{Y}{0}{n}.
\end{equation}
The estimation $\lsf{\hat{V}}{0}$, as opposed to the true value $\lsf{V}{0}$, depends on the estimation $\lsf{\hat{C}}{i}(s)$ and the simulation set.

\subsection{The LSM algorithm} \label{ssec:lsmintro} \noindent
The main difficulty in pricing Bermudan options with simulation methods lies in obtaining $\lsf{\hat{C}}{i}(s)$ (henceforth $\lsv{C}{i}$) from the simulated paths. This is primarily because the Monte Carlo path generation goes \textit{forward} in time, whereas the dynamic programming for pricing works \textit{backward} in time by construction.
\citet{longstaff2001val} obtain the estimate $\lsf{\hat{C}}{i}_\LSM(s)$ as the ordinary least squares (OLS) regression of the next path-wise option values $\lsv{Y}{i+1}$ on the current state $\lsv{X}{i}$: $$\lsf{\hat{C}}{i}_\LSM(s) = \lsf{X}{i}(s)\,\lsv{\beta}{i},$$
where $\lsv{\beta}{i}$ is a length-$M$ column vector of the regression coefficients.
Omitting the exercise time superscripts $[i]$ from $\lsv{X}{i}$ for simple notation, $\lsv{C}{i}_\LSM$ and $\lsv{\beta}{i}$ are
$$
\lsv{C}{i}_\LSM = \mat{X} \lsv{\beta}{i} = \lsv{H}{i}\lsv{Y}{i+1} \qtext{where}  \lsv{\beta}{i} =  \xtopxinv \mat{X}^\top \lsv{Y}{i+1}, \quad \lsv{H}{i} = \mat{X} \xtopxinv \mat{X}^\top,
$$
where $\lsv{H}{i}$ is the so-called hat matrix. Note that $\lsv{H}{i}$ depends on the current state, $\lsv{X}{i}$, not on the future information, $\lsv{Y}{i+1}$. Using (\ref{eq:backind1}), we inductively run the regression for $i=I-1, \cdots, 1$ until we obtain the option price $\lsf{\hat{V}}{0}_\LSM$.

To identify how look-ahead bias arises in the LSM algorithm, we focus on the exercise decision at time $t_i$ and state $s$. For this purpose, we consider only the simulations of size $N$ that have a path passing through the state $\lsvel{S}{i}{n}=s$ for a dummy path index $n$. Taking the expectation of (\ref{eq:backind1}) over such simulation sets, the option value from the LSM method (with the choice of basis functions) is
$$
\ExpMC[\lsf{\hat{V}}{i}_\LSM(s)] = \ExpMC[\lsvel{Y}{i}{n}] = \ExpMC[\,\mathcal{I}[\lsvel{C}{i}{n,\LSM} \ge \lsvel{Z}{i}{n}] \cdot (\lsvel{Y}{i+1}{n} - \lsvel{Z}{i}{n})\,|\, \lsvel{S}{i}{n}=s\,] + \lsf{Z}{i}(s).
$$
Ideally, the exercise decision, $\mathcal{I}[\lsvel{C}{i}{n, \LSM} \ge \lsf{Z}{i}(s)]$, and the continuation premium, $\lsvel{Y}{i+1}{n} - \lsvel{Z}{i}{n}$, should be independent because the former cannot take advantage of the future information of the simulation path. In the LSM method, however, $\lsvel{C}{i}{n, \LSM}$ depends on $\lsvel{Y}{i+1}{n}$ via $\lsv{C}{i} = \lsv{H}{i}\lsv{Y}{i+1}$. Therefore, the source of look-ahead bias is the covariance between the two terms:
\begin{equation} \label{eq:bias}
\lsf{B}{i}(s) = \Cov_\omega\left(\mathcal{I}[\lsvel{C}{i}{n, \LSM} \ge \lsvel{Z}{i}{n}],\,\lsvel{Y}{i+1}{n} - \lsvel{Z}{i}{n}\,|\, \lsvel{S}{i}{n}=s\,\right).
\end{equation}
Look-ahead bias is positive because $\lsvel{C}{i}{n, \LSM}$ is always biased toward $\lsvel{Y}{i+1}{n}$.

We can remove look-ahead bias by de-correlating $\lsv{C}{i}$ from $\lsv{Y}{i+1}$. One method is the standard technique of running an independent simulation set to estimate $\lsf{\hat{C}}{i}(s)$. Applying this type of method, say $\textsc{lsm}'$, to remove look-ahead bias, the option value from the method is suboptimal:
\begin{align*}
\ExpMC[\lsf{\hat{V}}{i}_{\LSM'}(s)] &= \ExpMC[\, \mathcal{I}[\,\lsvel{C}{i}{n,\LSM'} \ge \lsvel{Z}{i}{n}\,] \,|\, \lsvel{S}{i}{n}=s\,]\cdot \ExpMC[\,\lsvel{Y}{i+1}{n} - \lsvel{Z}{i}{n}\,|\, \lsvel{S}{i}{n}=s\,]+ \lsf{Z}{i}(s) \\
&= \lsf{p}{i}_{\LSM'}(s) \, \ExpMC[\,\lsf{Y}{i+1}\,|\, \lsvel{S}{i}{n}=s\,] + (1-\lsf{p}{i}_{\LSM'}(s))\lsf{Z}{i}(s) \\
&= \lsf{p}{i}_{\LSM'}(s)\, \lsf{C}{i}(s) + (1-\lsf{p}{i}_{\LSM'}(s))\lsf{Z}{i}(s) \\
&\le \max(\lsf{C}{i}(s), \lsf{Z}{i}(s)) = \lsf{V}{i}(s).
\end{align*}
Here, $\lsf{p}{i}_{\LSM'}(s)=\ExpMC[\, \mathcal{I}[\,\lsvel{C}{i}{n, \LSM'} \ge \lsvel{Z}{i}{n}\,] \,|\, \lsvel{S}{i}{n}=s\,]$ is the exercise probability at state $s$ averaged over repeated simulations. 

Our look-ahead bias expression is subtly different from that of \citet{fries2005foresight,fries2008foresight}. He defines it as the value of the option on the Monte Carlo error in the estimation of the continuation values:
$$ 
\lsf{B}{i}_\text{Fries}(s) = \Cov_\omega\left(\mathcal{I}[\lsvel{C}{i}{n,\LSM} \ge \lsvel{Z}{i}{n}],\,\lsvel{C}{i}{n,\LSM} - \lsvel{Z}{i}{n} \,|\, \lsvel{S}{i}{n}=s\right).
$$
We argue that this definition is inconsistent because it is based on the alternative backward induction (see footnote \ref{fn:alt}), even though \citet{fries2005foresight,fries2008foresight} claim to deal with the look-ahead bias in the LSM method. 

\subsection{The LOOLSM algorithm} \label{ssec:loolsm} 
\begin{figure}[!htb]
	\caption{\label{fig:loocv} Illustration of the look-ahead bias correction via LOOCV. The $x$-axis is the current state variable $S^{[i]}_n$, and the $y$-axis is the continuation premium $Y^{[i+1]}_n-Z^{[i]}_n$. There are three simulated paths: $n=1, 2$, and 3. The full regression ($\hat{y}=1+x$) under the LSM method indicates that path 2 should be \textit{continued} ($\hat{y}_2=1>0$). However, this is a look-ahead bias influenced by an outlier, path 2 ($y_2=4$). Based on the regression without path 2 ($\hat{y}'=-\frac23+\frac56 x$) under the LOOLSM method, it should be \textit{exercised} ($\hat{y}'_2=-\frac23<0$). The leverage scores, $h_n^{[i]}$, are $13/14$, $5/14$, and $10/14$.}
	\vspace{1ex} \centering
	\begin{tikzpicture}[x=1cm,y=0.7cm]
	\draw[step=2,gray,very thin,dotted] (-5,-5) grid (3,5);
	
	\draw [very thick, ->] (-5.5,0) -- (3.5,0) node [below] {$x=S^{[i]}_1$};
	\draw [very thick, ->] (0,-5) -- (0,5) node [above] {$\;y=Y^{[i+1]}_n-Z^{[i]}_n$};
	
	\draw [thick] (-5,-4) -- (3,4) node [right] {$\hat{y}=1+x$};
	\draw [very thick,dashed] (-5,-4.833333333) -- (3,1.833333333) node [right] {$\hat{y}'=-\frac23 + \frac56 x$};
	
	\path (-4,-4.02) coordinate (A);
	\path (0,4-0.02) coordinate (B);
	\path (2,1-0.02) coordinate (C);
	
	\draw (A) node[blue,fill,circle,scale=0.75]{};
	\draw (B) node[red,fill,rectangle,scale=0.9]{};
	\draw (C) node[blue,fill,circle,scale=0.75]{};
	
	\node at (A) [below right] {Path \textbf{1}: $(-4,-4)$};
	\node at (B) [left] {Path \textbf{2}: $(0,4)\;$};
	\node at (C) [below right]{Path \textbf{3}: $(2,1)$};
	
	\node at (0,1) {$-$};
	\node at (-1,1) {$\hat{y}_2 = 1$};
	
	\node at (0,-2/3) {$-$};
	\node at (1,-1) {$\hat{y}'_2 = -\frac23$};
	
	\node at (-1,1.75) {Continue};
	\node at (1,-1.75) {Exercise};
	\end{tikzpicture}	
\end{figure}
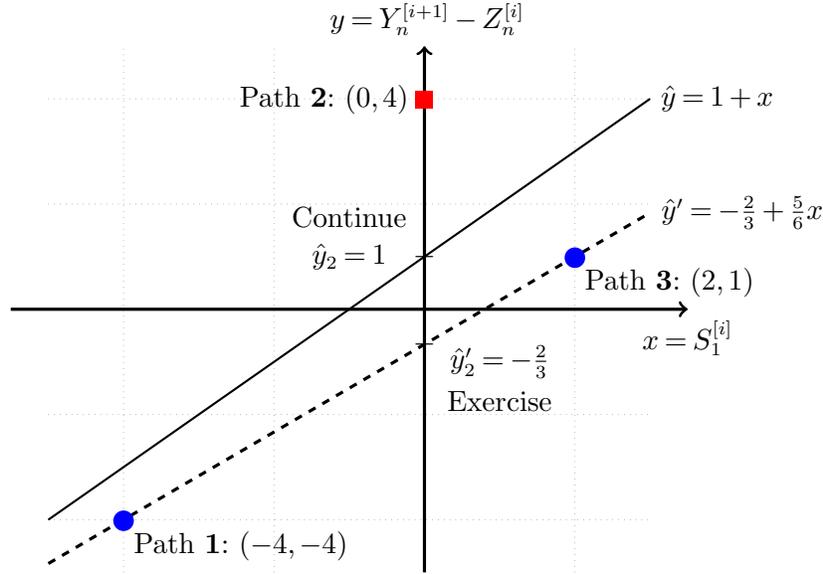

\noindent
We remove look-ahead bias in the LSM method simply by omitting each simulation path from the regression and making the exercise decision on the path from the self-excluded regression. The bias formulation (\ref{eq:bias}) is free from the correlation because we exclude $\lsvel{Y}{i+1}{n}$ from the estimation of $\lsvel{C}{i}{n}$. Figure~\ref{fig:loocv} illustrates this idea with a toy example with three simulation paths.

This idea is well known as LOOCV in statistical learning. This is a special type of the $k$-fold cross-validation method, where $k$ is equal to the number of data points $n$. We can analytically obtain the adjusted prediction values without running regressions $N$ times. We express the prediction error with the leave-one-out regression as a correction to that with the full regression~\citep[(7.64)]{hastie2009ESL}:\footnote{For derivation, see \url{https://robjhyndman.com/hyndsight/loocv-linear-models/}.}
$$
\lsv{Y}{i+1} - \lsv{C}{i}_\LOOLSM = \frac{\lsv{Y}{i+1} - \lsv{C}{i}_\LSM}{\vect{1}_N - \lsv{h}{i}},
$$
where $\vect{1}_N$ is the size-$N$ column vector of 1s, $\lsv{h}{i} = (\lsvel{h}{i}{n})$ is the diagonal vector of $\lsv{H}{i}$, and the arithmetic operations between vectors are element-wise. The diagonal element $\lsvel{h}{i}{n}$, measures the \textit{leverage} of the prediction $\lsvel{C}{i}{n}$ on the observation $\lsvel{Y}{i+1}{n}$; that is, $\lsvel{h}{i}{n} = \partial \lsvel{C}{i}{n} / \partial \lsvel{Y}{i+1}{n}$. The value is high when the observation point $\lsv{x}{i}_n$ is far enough away from the others that the regression is more likely fitted close to the observation (see the leverage values in the caption to Figure~\ref{fig:loocv}). It also satisfies\footnote{The lower bound $1/n$ occurs due to the intercept column in $\lsv{X}{i}$. We exclude the case where $\lsvel{h}{i}{n}=1$ because it happens only when $\lsv{X}{i}$ with the $n$-th observation removed is not full-rank. For the proof and equality condition, see \citet{mohammadi2016bounds}.} 
\begin{equation} \label{eq:h_bound}
\frac1n \le \lsvel{h}{i}{n} < 1 \qtext{and} \sum_{n=1}^N \lsvel{h}{i}{n} = \text{rank}(\lsv{X}{i}) = M.
\end{equation}
Note that the leave-one-out error is larger in magnitude than the original error because the full regression contains overfitting due to self-influence.

The LOOLSM method we propose is to use the corrected continuation value $\lsvel{C}{i}{n,\LOOLSM}$ from the LOOCV in the backward induction step (\ref{eq:backind1}):
\begin{equation} \label{eq:2.3}
\lsv{C}{i}_\LOOLSM = \lsv{C}{i}_\LSM - \frac{\lsv{h}{i} \cdot \lsv{e}{i}}{\vect{1}_N-\lsv{h}{i}}
\qtext{for} \lsv{e}{i} = \lsv{Y}{i+1} - \lsv{C}{i}_\LSM.
\end{equation}
We can compute the whole vector $\lsv{h}{i}$ as the row sum of the element-wise multiplication between $\mat{X}$ and $\mat{X}\xtopxinv$, which is straightforward from $\lsvel{h}{i}{n} = \vect{x}_n \xtopxinv \,\vect{x}_n^\top$. This is much more efficient than obtaining $\lsv{h}{i}$ from the full $\lsv{H}{i}$ matrix. As we must compute the transpose of $\mat{X} \xtopxinv$ for the full regression, we can obtain $\lsv{h}{i}$ with only $O(NM)$ additional operations. 

\section{Convergence rate of look-ahead bias} \label{sec:labias} \noindent
\subsection{Measuring look-ahead Bias} \label{ssec:measurelabis} \noindent
In this section, we analyze the convergence rate of look-ahead bias via the LOOLSM method. Given that the LOOCV correction removes the self-influence in the continuation price estimation, it is natural to define look-ahead bias as the difference between the LSM and LOOLSM prices:
\begin{equation} \label{eq:3.1}
\lsvel{B}{i}{n} = \lsvel{Y}{i}{n} - \lsvel{Y}{i}{n, \LOOLSM} \qtext{and} 
\hat{B} = \ExpN[\lsvel{B}{0}{n}] = \lsf{\hat{V}}{0}_\LSM - \lsf{\hat{V}}{0}_\LOOLSM,
\end{equation}
where $\lsvel{B}{i}{n}$ is the path-wise bias and $\hat{B}$ is the final bias in the option value at $t=0$. To keep the notation simple, we use $\lsvel{Y}{i}{n}$ instead of $\lsvel{Y}{i}{n,\LSM}$.
Measuring look-ahead bias with the LOOLSM method has two advantages over using the two-pass LSM method. First, we eliminate Monte Carlo error significantly because no extra randomness is required (see Table~\ref{tab:put2}). Moreover, we can mathematically analyze the convergence rate of look-ahead bias thanks to the analytic expression in \eqref{eq:2.3}.

There is a subtle difficulty in analyzing $\lsvel{B}{i}{n}$. The difference between $\lsvel{Y}{i}{n}$ and $\lsvel{Y}{i}{n, \LOOLSM}$ is easy to measure only when the LSM and LOOLSM methods use the same observations, $\lsv{Y}{i+1}=\lsv{Y}{i+1}_\LOOLSM$, for regression. However, this is only guaranteed at the first induction step $i=I-1$. From the next step, $\lsv{Y}{i+1}$ and $\lsv{Y}{i+1}_\LOOLSM$ start deviating. To resolve this difficulty, we introduce a modified LOOLSM method that substitutes $\lsv{Y}{i}$ for $\lsv{Y}{i}_\LOOLSM$ in the LOOLSM regression. 
\begin{equation} \label{eq:modified}
\begin{gathered}
\lsvel{Y}{i}{n} =  \mathcal{I}[\lsvel{C}{i}{n,\LSM} \ge \lsvel{Z}{i}{n}]\cdot (\lsvel{Y}{i+1}{n} - \lsvel{Z}{i}{n}) + \lsvel{Z}{i}{n}, \\
\lsvel{Y}{i}{n,\LOOLSM} =  \mathcal{I}[\lsvel{C}{i}{n,\LOOLSM} \ge \lsvel{Z}{i}{n}]\cdot (\lsvel{Y}{i+1}{n, \LOOLSM} - \lsvel{Z}{i}{n}) + \lsvel{Z}{i}{n}, \\
\lsvel{C}{i}{n, \LOOLSM} = \lsvel{C}{i}{n, \LSM} - (\lsvel{Y}{i+1}{n} - \lsvel{C}{i}{n,\LSM})\, \lsvel{h}{i}{n} / (1-\lsvel{h}{i}{n})
\end{gathered}
\end{equation}
We numerically verify that the impact of this modification is negligible in pricing because the substitution affects only the estimated continuation value. Therefore, we assume that the LOOLSM price in (\ref{eq:3.1}) is measured with the modified method throughout this section.

\subsection{Main result} \label{ssec:mainres} \noindent
Before stating the main result, we first build an intuition for the convergence rate of look-ahead bias. Suppose the $n$-th path is an outlier in the LSM regression such that the sample point $\lsvel{Y}{i+1}{n}$ is much bigger than the prediction $\lsvel{C}{i}{n, \LSM}$.\footnote{By symmetry, one can also assume that $\lsvel{Y}{i+1}{n}$ is much smaller than $\lsvel{C}{i}{n, \LSM}$.} Then, look-ahead bias inverts the exercise decision when $ \lsvel{C}{i}{n, \LOOLSM} < \lsvel{Z}{i}{n} \le \lsvel{C}{i}{n, \LSM}$. In Lemma~\ref{lemma:A.1} in \ref{apdx:deriveasympt}, we will show that this is equivalent to
\begin{equation}\label{eq:2.4}
0 \le \lsvel{C}{i}{n, \LSM} - \lsvel{Z}{i}{n} < \lsvel{h}{i}{n} (\lsvel{Y}{i+1}{n} - \lsvel{Z}{i}{n}).
\end{equation}
We can guess that the probability for the above events would decay at the rate of $M/N$. This is because, as the simulation size $N$ grows larger, $\lsvel{h}{i}{n}$ becomes smaller as $\ExpN [\lsvel{h}{i}{n}] = M/N$ from \eqref{eq:h_bound} while the size of the other terms remains the same. Indeed, this turns out to be the case, as we discuss below in detail.

The main results rely on two technical assumptions common in studies analyzing the convergence of the LSM algorithm~\citep{clement2002anal,stentoft2004conv}. First, we work only with realistic payoff functions that grow moderately and well-defined option prices. This is a minimal condition from a practical standpoint.
\begin{assumption}\label{assumption:1}
	The payout functions, $\lsf{Z}{i}(s)$, are in $L^2$. 
\end{assumption}

The second assumption deals with the complication in pricing that arises when the option and continuation values are arbitrarily close with a non-negligible probability. This outcome might lead to a wrong exercise decision at the limit, and the LSM algorithm fails to converge to the true price; see~\citet{stentoft2004conv} for example. Henceforth, we assume that the continuation value is almost certainly different from the exercise value. 

\begin{assumption}
	\label{assumption:2}
	Fix an ordered set of countable basis functions $\{f_m(s):\, m=0,1,\cdots \}$ in the $L^2$ space. Let $\lsvel{C}{i}{M}(s)$ be the continuation value obtained from the LSM method with the first $M$ basis functions at the limit as $N \rightarrow \infty$, such that $\lsvel{C}{i}{M}(s) \rightarrow \lsf{C}{i}(s)$ as $M\rightarrow \infty$. Furthermore, let 
	$$\lsvel{P}{i}{M}(c) = \mathbb{P}[\,|\lsvel{C}{i}{M}(\lsf{S}{i}) - \lsvel{Z}{i}{}(\lsf{S}{i})| \le c \, ]$$ 
	be the probability of the absolute continuation premium not exceeding $c$.
	Then, we assume that 
	$$ \lim_{c \rightarrow 0}\lsvel{P}{i}{M}(c) = 0 \qtext{for all} M=1,2,\cdots.$$
\end{assumption}

The following theorem is the main result of this section. In analyzing the convergence rate, we regard any derived quantity (e.g., $\hat{B}$) as a random variable, and examine how its expected value behaves as $N$ increases. 
\begin{restatable}{theorem}{thmepsilon}
	\label{theorem:3.1}
	The following hold under Assumptions~\ref{assumption:1} and \ref{assumption:2}.
	\begin{enumerate}[(i)] \normalfont
		\item $\lsvel{B}{i}{n} \sim O_p(M/N)$.
		\item For any given $\varepsilon>0$, there exists $r_\varepsilon>0$ such that the expected look-ahead bias satisfies $\ExpMC[\hat{B}] \le \varepsilon + r_\varepsilon M/N$.
		\item $\hat{B}$ converges to zero in probability.
	\end{enumerate}
	Here, the probabilistic asymptotic notation $O_p$ is defined in the probability space of all possible simulation runs of size $N$. The subscript for path $n$ in (i) is a dummy index because the Monte Carlo paths are drawn independently.
\end{restatable}

We prove Theorem~\ref{theorem:3.1} in \ref{apdx:deriveasympt}. Two cases are treated separately in the proof: (a) the contributions to look-ahead bias near the exercise boundary and when the tails of the asset distributions can be made arbitrarily small, and (b) the probability of look-ahead bias occurring elsewhere can be bounded by a constant multiple of leverage $\lsvel{h}{i}{n}$ with expected value $M/N$. Since any realistic bias is controlled by (b), its expected value decays at the rate of a constant multiple of $M/N$ at least. Indeed, we report a strong linear relationship between look-ahead bias and $M/N$ with a few examples; see Figures~\ref{fig:put}, \ref{fig:bestof}, and \ref{fig:basket}. Although Theorem~\ref{theorem:3.1} does not guarantee linearity, we believe this is a direct consequence of $\lsvel{h}{i}{n}$ primarily determining the convergence rate.

\section{Numerical results} \label{sec:num} \noindent
\subsection{Overview of experiments} \noindent
We price four Bermudan option cases to compare the LSM and LOOLSM methods. We present them in increasing order of the number of underlying assets: single-stock put options, best-of options on two assets, basket options on four assets, and cancellable exotic interest swaps under the LIBOR market model. Therefore, the number of regressors, $M$, also increases in general, given the same polynomial orders to include. 

We run $n_\textsc{mc}$ sets of simulations with $N$ paths each and use the following three estimators for comparison:
\begin{itemize}
	\item \textbf{LSM}: the one-pass (i.e., in-sample) LSM estimator.
	\item \textbf{LSM-2}: the two-pass (i.e., out-of-sample) LSM estimator.
	We apply the exercise policy computed from an extra set of $N$ paths to the payoff valuation with the original simulation set.
	\item \textbf{LOOLSM}: the LOOLSM estimator.
\end{itemize}
Using the same $N$ simulation paths for the payoff valuation across the three methods works as a control to reduce the variability of the measured bias (i.e., the price difference between methods). We reduce the variance by using the antithetic random variate ($N/2+N/2$).\footnote{Antithetic variate is applied across paths (i.e., $N/2$ paths + $N/2$ paths), not within each path.} We also vary $M$ by selecting different basis sets. From the results of $n_\textsc{mc}$ independent simulation sets, we obtain the mean and standard deviation of the option price. If the exact option value $\lsf{V}{0}$ is available, then we report the price offset from $\lsf{V}{0}$:
$$ 
\quad\text{Price Offset} = \ExpMC[\lsf{\hat{V}}{0}] - \lsf{V}{0}, 
$$
where $\lsf{\hat{V}}{0}$ is the price estimate from each simulation. Otherwise, we report the price $\ExpMC[\lsf{\hat{V}}{0}]$. We implemented the numerical experiments in Python (Ver. 3.7, 64-bit) on a computer running Windows 10 with an Intel core i7 1.9 GHz CPU and 16 GB RAM.

\subsection{Bermudan options under the Black-Scholes model} \label{ssec:put} \noindent
The underlying asset prices $S_j(t)$ of the three examples in this section follow geometric Brownian motions:
$$ \frac{dS_j(t)}{S_j(t)} = (r-q_j)dt + \sigma_j\, dW_j(t),
$$
where $r$ is the risk-free rate, $q_j$ is the dividend yield, $\sigma_j$ is the volatility, and the $W_j(t)$'s are the standard Brownian motions correlated by
$dW_j(t)\, dW_{j'}(t) = \rho_{jj'}\,dt$ ($\rho_{jj}=1$).
The choice of geometric Brownian motion for the price dynamics has several advantages and does not oversimplify the problem. It is easy to implement because an exact simulation is possible. The geometric Brownian motion is a standard choice in the literature and we can take advantage of the exact Bermudan option prices reported previously. 
The three examples in this section are the Bermudan options whose payouts are determined entirely at the time of exercise and are always non-negative. We take advantage of these properties in the implementation of the methods.\footnote{First, we include the payout function as a regressor, $f_1(s) = \lsf{Z}{i}(s)$. The payout is an important regressor improving the optimality of the exercise decision, as \citet{glasserman2003mc_ch8} shows. Including the payout also eliminates a specification issue in regression. In an alternative LSM implementation, one may regress the continuation premium, $\lsv{Y}{i+1} - \lsv{Z}{i}$ (instead of $\lsv{Y}{i+1}$), to estimate $\lsv{C}{i} - \lsv{Z}{i}$ (instead of $\lsv{C}{i}$). While it is difficult to determine the superior approach, they become identical when we include $\lsf{Z}{i}(s)$ as a basis function. Second, following \citet{beveridge2013ppi}, we do not exercise the option when $\lsvel{Z}{i}{n}=0$, even if $\lsvel{C}{i}{n}<\lsvel{Z}{i}{n}$. The negative continuation value is an artifact caused by simulation noise or imperfect basis functions. It is always optimal to continue the option since the future payout is non-negative. On top of the exercise decision override, \citet{longstaff2001val} even suggest running the regressions with the in-the-money paths only; that is, $\{n:\; \lsvel{Z}{i}{n}>0\}$. As \citet{glasserman2003mc_ch8} reports that the result can be inferior in some cases, we use all simulation paths in the numerical experiments except Case 1. We use in-the-money paths only in Case 1 because the purpose of Case 1 is to reproduce the result of \citet{longstaff2001val}.
}

For each example in this section, we run two experiments. The first experiment is to ensure that the LOOLSM method eliminates look-ahead bias similarly to the LSM-2 method. We run $n_\textsc{mc}=100$ sets of simulations with $N=4\times 10^4$ paths and price options with the three methods. Additionally, we price the corresponding European options using the Monte Carlo method with the same sets of paths.

The second experiment is to validate the convergence rate of the bias in Theorem~\ref{theorem:3.1}. We run the LSM and LOOLSM methods with varying $N$ and $M$. We first generate a pool of $720\times 10^4$ Monte Carlo paths and split them into groups of $N=0.5,\,1.5,\,3,\,6,\,9,\,12,\,18,$ and $36 \; \times \; 10^{4}$ paths. Therefore, the generated groups comprise $n_\textsc{mc} = 720\times 10^{4} / N$ ($=1440,\cdots, 20$) Monte Carlo runs, and we compute the price offset and standard deviation from the $n_\textsc{mc}$ prices.
By varying $N$ within the same path pool, we control the Monte Carlo variance as much as possible and make the simulation size $N$ the most important factor to measure look-ahead bias. Simultaneously, we vary the number of regressors ($M$) by including polynomials of higher terms. We thus measure look-ahead bias as a function of $M/N$.

\noindent
\textbf{Case 1: Single-stock Put Option}. We start with Bermudan put options on a single stock
$$ \lsf{Z}{i}(s) = e^{-rt_i} \max(K-s_1,0),
$$
with the parameter set tested in \citet{longstaff2001val}:
$$ K = 40,\; \sigma_1 = 40\%,\; r=6\%,\; q_1=0\%,\; t_i = \frac{i}{50}, \text{ and } I=50 \; (T=1).
$$
We adopt the pricing details as much as possible from \citet{longstaff2001val} to replicate the result. We use in-the-money paths only for regression. For the regressors, we use the set of weighted Laguerre polynomials as in \citet{longstaff2001val}:
$$ \lsf{X}{i}(s) = \left(1,\, e^{-\frac{s_1}{2}}, \, e^{-\frac{s_1}{2}}(1-s_1),\, e^{-\frac{s_1}{2}}\left(1-2s_1+\frac{s_1^2}{2}\right),\;\cdots\right).
$$
The first $M=4$ functions (the first three weighted Laguerre polynomials) are used for the first experiment, and $M=4$, 6, and 8 functions are used for the second experiment. We also use the exact option prices reported in \citet{longstaff2001val}.

Table~\ref{tab:put1} reports the result of the first experiment. As expected, the LOOLSM and LSM-2 prices are similar and are slightly lower than the true prices. On the contrary, the LSM prices are higher than the true prices, indicating look-ahead bias. This result implies that LOOLSM removes look-ahead bias, although the size is small. Table~\ref{tab:put2} reports its mean and standard deviation separately to show the look-ahead bias's statistical significance. The bias measured with the LOOLSM method has much less deviation than that measured with LSM-2 because the LOOLSM method requires no extra simulation, whereas LSM-2 needs another independent simulation set. When measured as LSM $-$ LOOLSM, the average look-ahead bias is at least twice as big as the standard deviation. Therefore, the look-ahead bias is statistically significant. 

\begin{figure}[!htb]
	\caption{\label{fig:put} The price offset (top) and look-ahead bias (bottom) as functions of $M/N$ for the single-stock put option with $S_1(0)=36$ (Case 1). At the top, given the fixed values of $M$ and $N$, the higher value corresponds to the LSM method and the lower one to the LOOLSM method.}
	\vspace{1ex} \centering
	\includegraphics[width=0.55\linewidth]{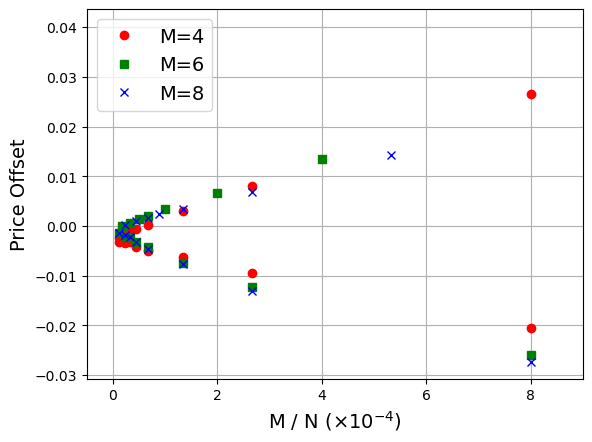} \\ \vspace{1ex}
	\includegraphics[width=0.55\linewidth]{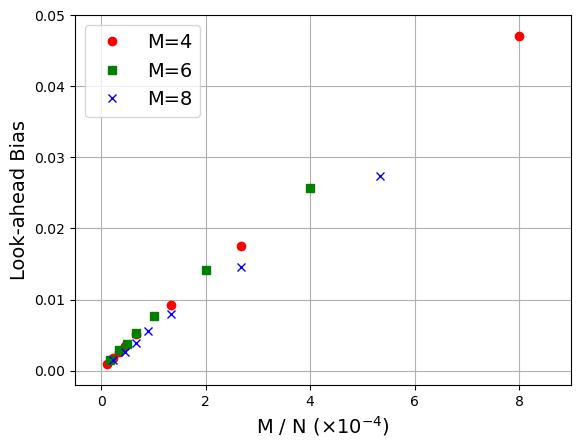}
\end{figure}

Figure~\ref{fig:put} shows the result of the second experiment. The top plot shows the price offset of the LSM and LOOLSM methods as a function of $M/N$ for varying $M$ and $N$ values. It demonstrates how the LSM and LOOLSM prices converge as $N$ increases for a fixed $M$. The LSM price converges from above, and the LOOLSM converges from below, indicating that the LOOLSM price is lower-biased than the convergent value for a given $M$. The bottom plot shows the look-ahead bias as a function of $M/N$. Notably, the data from the three $M$ values form a clear linear pattern, confirming the convergence rate in Theorem~\ref{theorem:3.1}. While the figure illustrates one specific option ($S_1(0)=36$), the other options in the case exhibit the same pattern.

Our conclusion opposes the analysis in \citet{longstaff2001val}. In Table~2 and related comments in the paper, they compare the ``in sample'' (i.e., LSM) and ``out of sample'' (i.e., LSM2) prices obtained from $N=100,000$ paths and $M=4$ regressors, and concludes that the difference (i.e., look-ahead bias) is negligible. This is indeed consistent with our results; with $M/N=0.4\times 10^{-4}$, the bias is very small. However, this should not be generalized to any LSM method application, as the bias can grow larger as $M/N$ varies. The LSM price might rise above the true price when larger basis sets are used with a smaller number of paths.


\begin{table}[!htb]
	\caption{\label{tab:put1} 
		Results for the single-stock Bermudan put options (Case 1). We use $N=40,000$, $M=4$. The ``Exact'' columns report the true option prices, while the other columns report the price offset and standard deviation from the $n_\textsc{mc}=100$ simulation results. All values are rounded to three decimal places.} \centering
	\vspace{1ex}
	\begin{tabular}{|c||c|c|c|c||c|c|}
		\hline
		& \multicolumn{4}{c||}{Bermudan} & \multicolumn{2}{c|}{European} \\ \hline
		$S_1(0)$ & Exact & LSM & LSM-2 & LOOLSM & Exact & MC \\ \hline
36 & 7.101 & 0.003 $\pm$ 0.014 & -0.004 $\pm$ 0.014 & -0.004 $\pm$ 0.015 & 6.7114 & 0.0006 $\pm$ 0.0177 \\
38 & 6.148 & 0.002 $\pm$ 0.012 & -0.006 $\pm$ 0.013 & -0.005 $\pm$ 0.012 & 5.8343 & 0.0007 $\pm$ 0.0203 \\
40 & 5.312 & 0.004 $\pm$ 0.013 & -0.003 $\pm$ 0.013 & -0.003 $\pm$ 0.012 & 5.0596 & 0.0008 $\pm$ 0.0222 \\
42 & 4.582 & 0.004 $\pm$ 0.012 & -0.003 $\pm$ 0.013 & -0.003 $\pm$ 0.012 & 4.3787 & 0.0009 $\pm$ 0.0232 \\
44 & 3.948 & 0.003 $\pm$ 0.014 & -0.004 $\pm$ 0.014 & -0.004 $\pm$ 0.015 & 3.7828 & 0.0008 $\pm$ 0.0235 \\
		\hline
	\end{tabular}
\end{table}

\begin{table}[!htb]
	\caption{\label{tab:put2} Results for the first experiment on the single-stock Bermudan put options in \S~\ref{ssec:put}. The columns report the difference between the price of each method and the LSM price, and its error estimate.} \centering 
	\vspace{1ex}
	\begin{tabular}{|c||c|c|}
		\hline
		$S_1(0)$ & LSM $-$ LSM-2 & LSM $-$ LOOLSM \\ \hline
36 & 0.0069 $\pm$ 0.0068 & 0.0070 $\pm$ 0.0027 \\
38 & 0.0077 $\pm$ 0.0078 & 0.0073 $\pm$ 0.0027 \\
40 & 0.0066 $\pm$ 0.0067 & 0.0069 $\pm$ 0.0028 \\
42 & 0.0069 $\pm$ 0.0065 & 0.0069 $\pm$ 0.0023 \\
44 & 0.0065 $\pm$ 0.0057 & 0.0068 $\pm$ 0.0024 \\
		\hline
	\end{tabular}
\end{table}

\noindent
\textbf{Case 2: Best-of Option on Two Assets}.
We price best-of (or rainbow) call options on two assets:
$$ 
\lsf{Z}{i}(s) = e^{-rt_i} \max(\max(s_1,s_2)-K,0),
$$
with the parameter set tested by \citet{glasserman2003mc_ch8} and \citet{andersen2004primal}:
$$ 
K = 100,\; \sigma_j = 20\%,\; r=5\%,\; q_j=10\%,\; \rho_{j\neq j'}=0,\; t_i = \frac{i}{3}, \text{ and } I=9 \; (T=3).
$$
The options are priced with three initial asset prices, $S_1(0) = S_2(0) = 90,\, 100$, and 110. We use the following basis functions (M=11) for the first experiment:
$$
\lsf{X}{i}(s) = (1,\; \lsf{Z}{i}(s),\; s_1,\; s_2,\; s_1^2,\; s_1s_2,\; s_2^2,\; s_1^3, \;s_1^2s_2,\; s_1s_2^2,\; s_2^3).
$$
For the second experiment, we use $M=4$, 7, and 11, which correspond to the linear, quadratic, and cubic polynomial terms, respectively. We use the exact Bermudan option prices from \citet{andersen2004primal} and compute the exact European option prices from the analytic solutions expressed in terms of the bivariate cumulative normal distribution~\citep{rubinstein1994some}. 

\begin{figure}[!htb]
	\caption{\label{fig:bestof} The price offset (top) and look-ahead bias (bottom) as functions of $M/N$ for the best-of call option with $S_j(0)=100$ (Case 2). At the top, given the fixed values of $M$ and $N$, the higher value corresponds to the LSM method and the lower one to the LOOLSM method.}
	\vspace{1ex} \centering
	\includegraphics[width=0.55\linewidth]{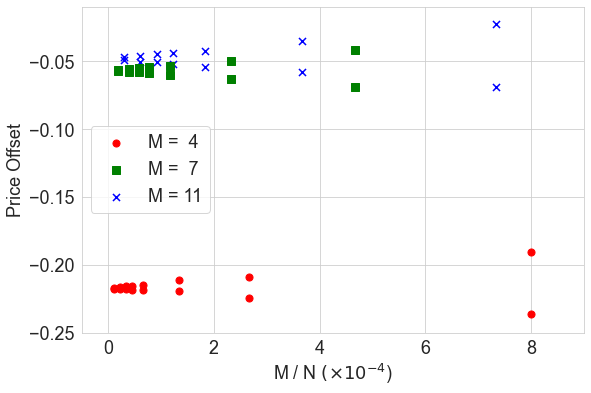} \\ \vspace{1ex}
	\includegraphics[width=0.55\linewidth]{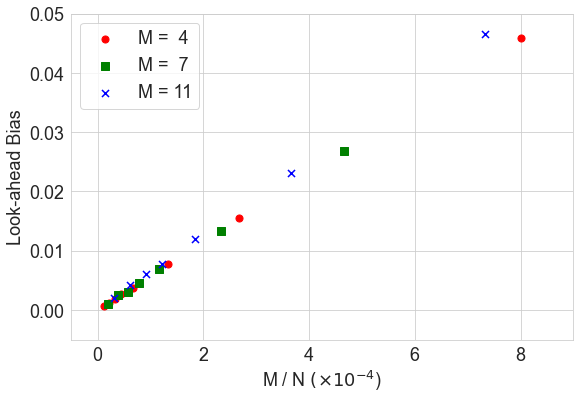}
\end{figure}

Table~\ref{tab:bestof} and Figure~\ref{fig:bestof} show the results. Look-ahead bias in the LSM method becomes more pronounced, whereas the LSM price is still lower than the true price, primarily because the exercise boundary of the best-of option is highly non-linear, as \citet{glasserman2003mc_ch8} observes. As depicted in Figure~\ref{fig:bestof}, suboptimal bias quickly decreases as $M$ increases. Nevertheless, look-ahead bias is clearly proportional to $M/N$, regardless of the suboptimality level.

\begin{table}[!htb]
	\caption{\label{tab:bestof} Results for the best-of Bermudan call options (Case 2). We use $N=40,000$ and $M=11$. The ``Exact'' columns report the true option prices, while the other columns report the price offset and standard deviation from $n_\textsc{mc}=100$ simulation results. All values are rounded to three decimal places.} \centering
	\vspace{1ex} 
	\begin{tabular}{|c||c|c|c|c||c|c|}
		\hline
		& \multicolumn{4}{c||}{Bermudan} & \multicolumn{2}{c|}{European} \\ \hline
		$S_j(0)$ & Exact & LSM & LSM-2 & LOOLSM & Exact & MC \\ \hline
		\, 90 & \, 8.075 & -0.020 $\pm$ 0.055 & -0.036 $\pm$ 0.056 & -0.035 $\pm$ 0.054 & \, 6.655 & 0.011 $\pm$ 0.062 \\
		100 & 13.902 & -0.036 $\pm$ 0.060 & -0.052 $\pm$ 0.062 & -0.054 $\pm$ 0.058 & 11.196 & 0.011 $\pm$ 0.078 \\
		110 & 21.345 & -0.040 $\pm$ 0.065 & -0.062 $\pm$ 0.068 & -0.059 $\pm$ 0.064 & 16.929 & 0.013 $\pm$ 0.096 \\
		\hline
	\end{tabular}
\end{table}

\noindent \textbf{Case 3: Basket Option on Four Assets}. \label{ssec:basket}
Next, we price Bermudan call options on a basket of four stocks:
$$ 
\lsf{Z}{i}(s) = e^{-rt_i} \max\left(\frac{s_1+s_2+s_3+s_4}{4}-K,0\right).
$$
with the parameter set tested by \citet{wilmott_basket} and \citet{choi2018sumbsm} in the context of the European payoff,
$$ 
S_j(0) = 100,\; \sigma_j = 40\%,\; r=q_j=0,\; \rho_{j\neq j'}=0.5,\; t_i = \frac{i}{2}, \text{ and } I=10 \;(T=5).
$$
The options are priced for a range of strikes, $K=60, 80, 100, 120$, and 140. Because the underlying assets do not pay dividends, it is optimal not to exercise the option until maturity; hence, the Bermudan option price equals the European price. Therefore, we use the European basket option price in \citet{choi2018sumbsm} for the exact Bermudan option prices. For the regressors, we use polynomials up to degree 2 ($M=16$) for the first experiment:
$$ 
\lsf{X}{i}(s) = (1,\,\lsf{Z}{i}(s),\, s_j,\cdots,\, s_j^2,\cdots,\, s_js_{j'},\cdots) \qtext{for} 1\le j<j'\le 4,
$$
and the subsets $M=6, 10$, and 16 for the second experiment.

\begin{figure}[!htb]
	\caption{\label{fig:basket} The price offset (top) and look-ahead bias (bottom) as functions of $M/N$ for the four-asset basket option with $K=100$ (Case 3). At the top, given the fixed values of $M$ and $N$, the higher value corresponds to the LSM method and the lower one to the LOOLSM method.}
	\vspace{1ex} \centering
	\includegraphics[width=0.55\linewidth]{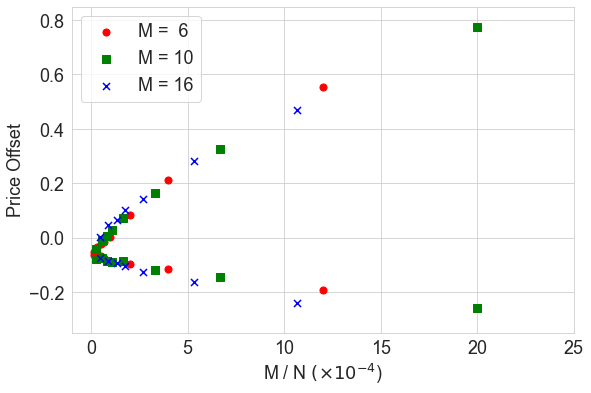} \\ \vspace{1ex}
	\includegraphics[width=0.55\linewidth]{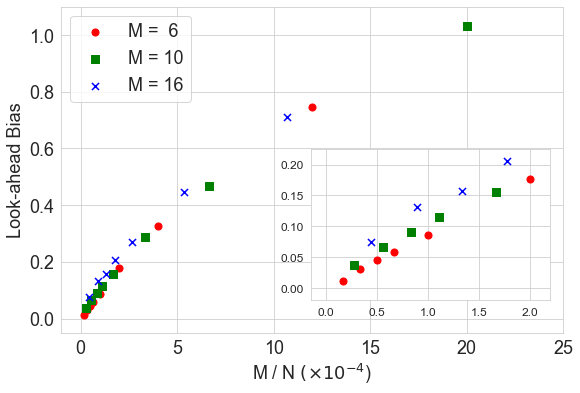}
\end{figure}

Table~\ref{tab:basket} and Figure~\ref{fig:basket} report the results. In this example, the LSM method noticeably overprices the option for all strike prices, whereas the LOOLSM and LSM-2 prices are consistently low-biased. Unlike the best-of option case, the suboptimal level is unchanged for increasing $M$ because the payoff function is a linear combination of the asset prices; therefore only the linear basis functions ($M=6$) capture the exercise boundary accurately. The look-ahead biases for the different $M$'s collapse into a function of $M/N$, although linear convergence clearly appears when $M/N$ is very small (see the inset of the bottom plot of Figure~\ref{fig:basket}).

\begin{table}[!htb]
	\caption{\label{tab:basket} Results for the four-asset basket options (Case 3). We use $N=40,000$ and $M=16$. The ``Exact'' columns report the true option prices, while the other columns report the price offset and standard deviation from $n_\textsc{mc}=100$ simulation results. All values are rounded to three decimal places.} 
	\centering
	\vspace{1ex}
	\begin{tabular}{|c||c|c|c|c||c|}
		\hline
		$K$ & Exact & LSM & LSM-2 & LOOLSM & European \\ \hline
		\, 60 & 47.481 & 0.233 $\pm$ 0.223 & -0.205 $\pm$ 0.213 & -0.209 $\pm$ 0.196 & 0.012 $\pm$ 0.309 \\
		\, 80 & 36.352 & 0.230 $\pm$ 0.255 & -0.174 $\pm$ 0.244 & -0.158 $\pm$ 0.235 & 0.012 $\pm$ 0.316 \\
		100 & 28.007 & 0.235 $\pm$ 0.237 & -0.117 $\pm$ 0.238 & -0.109 $\pm$ 0.231 & 0.012 $\pm$ 0.309 \\
		120 & 21.763 & 0.226 $\pm$ 0.236 & -0.084 $\pm$ 0.245 & -0.080 $\pm$ 0.229 & 0.013 $\pm$ 0.293 \\
		140 & 17.066 & 0.213 $\pm$ 0.224 & -0.086 $\pm$ 0.222 & -0.075 $\pm$ 0.223 & 0.015 $\pm$ 0.275 \\
		\hline
	\end{tabular}
\end{table}

\subsection{Cancellable exotic interest rate swap under the LIBOR market model} \label{ssec:cmsra} \noindent
Finally, we apply the LOOLSM method to a cancellable exotic interest rate derivative under the LIBOR market model~\citep{brace1997bgm,jamshidian1997libor}. This last example differs from the previous three examples in that it is closer to the structured product traded in the market and has a much higher computational cost than the previous examples. The exact option price is not available. We use this example to demonstrate the computational advantage of the LOOLSM algorithm. 

We briefly introduce the LIBOR market model first before introducing the payout. Let $P(t,T)$ denote the time $t$ price of the zero-coupon bond paying \$1 at $T$. For a set of equally spaced dates, $T_j = j\delta$ for a tenor $\delta$, the forward rate between $T_j$ and $T_{j+1}$ seen at time $t\le T_j$ is given by
$$
F_j(t) = \frac{1}{\delta}\left(\frac{P(t,T_j)}{P(t,T_{j+1})} - 1\right).
$$
We also denote the \textit{spot} rate by $F_j = F_j(T_j)$.
The LIBOR market model evolves $\{F_j(t)\}$, and then yields the discount curve $P(t,\cdot)$. Among the various model specifications, we follow the displaced-diffusion stochastic volatility implementation of \citet{joshi2003lmmsv}; the forward rates follow displaced geometric Brownian motions
$$
\frac{d\, F_j(t)}{F_j(t) + \alpha} = \mu_j(t)\, dt + \sigma_j(t)\, dW_j(t) \qtext{for} 0\le t\le T_j, 
$$
where the $W_j(t)$s are the correlated standard Brownian motions, and the volatility takes the time-homogeneous form
$$ \sigma_j(t) = \big(a+b(T_j-t)\big) e^{-c(T_j-t)} + d \qtext{for} 0\le t\le T_j.
$$
This \textit{abcd} volatility structure is popular in the literature~\citep{joshi2014eff,beveridge2013ppi}. \citet{joshi2003lmmsv} further makes volatility stochastic by letting $a$, $b$, $\log c$, and $\log d$ evolve over time following the Ornstein--Uhlenbeck process
\begin{equation} \label{eq:abcd_OU}
d\, h(t) = \lambda (\,h_\infty - h(t)) + \sigma_h dW_h(t),
\end{equation} 
where $W_h(t)$ is a standard Brownian motion independent of the $W_j(t)$s. Displaced diffusion and stochastic volatility enable the model to exhibit the swaption volatility smile observed in the market. We choose a spot measure where the numeraire asset is a discretely compounded money market account with \$1 invested at $t=0$. The value of the numeraire asset at $t=T_j$ is 
$$ P^*_j = \prod_{k=0}^{j-1} \left( 1 + \delta F_j\right)
$$
The drift, $\mu_j(t)$, is determined by the arbitrage condition depending on the numeraire choice. The predictor--corrector method~\citep{hunter2001drift} provides an efficient approximation of the integrated drift required for the simulation of $F_j(t)$. 

\noindent \textbf{Case 4: Cancellable CMS Spread and LIBOR Range Accrual.} Using the LIBOR market model, we price a callable structure note with an exotic coupon rate. In the equivalent swap form, the note issuer (option buyer) pays an exotic coupon with annual rate $R_j$ and the investor (option seller) pays the market rate $F_{j-1}$ at the end of each period, $t=T_j$. The exotic coupon is paid only when the spread of two constant maturity swaps (CMS) and LIBOR rates are within certain ranges. Specifically, we assume
$$ R_j =
\begin{cases} 
0.095 \qquad\qquad\qtext{in the first one year} (j=1, \ldots, 1/\delta) \\
0.095\cdot \mathcal{I}[\,\swap{2}_j \le \swap{10}_j\,] \cdot \mathcal{I}[\,0\le F_j\le 0.03\,] \quad\text{afterwards},
\end{cases}
$$
where $\swap{n}_j$ is the $n$-year swap rates implied from the forward rates at the time of cashflow exchange, $\{F_k(T_j): k\ge j\}$. The two conditions embedded in $R_j$ have been popular among investors since the financial crisis in 2008; the inversion of the swap curve (i.e., $\swap{2}_j > \swap{10}_j$) is historically rare, and it is expected that the Federal Reserve will maintain a low realized short-term rate (i.e., $F_j$). However, the risk-neutral probability of the conditions implied from the option market is lower. Therefore, the coupon rate (i.e., 9.5\%) can be set high to balance the present values of the two parties. While most market trades use daily range accrual to mitigate the fixing risk, our example uses a single observation per coupon period to simplify pricing.

We assume that the forward rate tenor is six months ($\delta = 0.5$) and that the swap matures in 20 years. Therefore, we simulate 60 forward rates, $\{F_j(t): 1\le j\le 60\}$, until $t=20$. The issuer can cancel the swap every year at $t_i = T_{2i}=i$ for $i= 1, \ldots, 20$ ($I=20$). Following the market convention, cancellation does not apply to the cashflow exchange at the same time. To price the trade in a Bermudan option form, we decompose the cancellable swap into two trades: the (non-cancellable) underlying swap and the Bermudan swaption, where the holder has the right to enter the swap from $t_i$ to $t=20$. The payout of the Bermudan swaption to the option holder is 
$$\lsf{Z}{i} = \sum_{j=2i+1}^{2I} \frac{\delta}{P^*_j} (R_j - F_{j-1}) \quad (\lsf{Z}{I} = 0).
$$
We obtain the price of the cancellable swap as the sum of the prices for the two trades accordingly. 

\begin{table}[!htb]
	\caption{The Ornstein--Uhlenbeck parameters of the stochastic \textit{abcd} volatility for cancellable exotic interest rate swap (Case 4). The parameter values are from \citet[Table 4]{joshi2003lmmsv}} \label{tab:abcd}
	\centering
	\vspace{1ex} 
	\begin{tabular}{|c|ccc|} \hline
		$h(t)$ & $h(0)\, (=h_\infty)$ & $\sigma_h$ & $\lambda$ \\ \hline
		$a$ & $-0.020$ & 0.05 & 0.5 \\
		$b$ & \hspace{1.5ex}0.108 & 0.1 & 0.3 \\
		$\log c$ & $\log(0.800)$ & 0.1 & 0.5 \\
		$\log d$ & $\log(0.114)$ & 0.2 & 0.4268 \\ 
		\hline
	\end{tabular}
\end{table}

We use the model parameters in \citet[\S~8.4]{joshi2003lmmsv} for the simulation. The displacement is $\alpha = 0.025$ and the correlation between forward rates decays exponentially, $dW_j(t)dW_{j'}(t) = e^{-\theta\delta|j-j'|}\,dt$ with $\theta=0.1$. The Ornstein--Uhlenbeck parameters for the \textit{abcd} volatility are given in Table~\ref{tab:abcd}. See \citet[\S~8]{joshi2003lmmsv} for the swaption volatilities implied from the parameter set. We simulate the stochastic volatility with the Euler scheme with time step, $\Delta t = 0.25$. The initial forward rates are $F_j(0) = 0.045 - 0.0425\, e^{-j\delta/4}$ such that $F_j$ increases from $0.25\%$ to $4.5\%$ as $j$ increases. 

Because the payout $\lsf{Z}{i}$ is not determined at the time of exercise, we require a different regression implementation from the previous cases. The basis function does not include $\lsf{Z}{i}$. We apply the regression to the continuation premium (or penalty), $\lsv{Y}{i+1} - \lsv{Z}{i}$ instead.\footnote{In general, we can separately regress $\lsv{Y}{i+1}$ and $\lsv{Z}{i}$ \citep{piterbarg2003guideCLE}.} For the basis set $\lsf{X}{i}$, we consider two groups of variables: (i) the six variables related to interest rates, $F_{2i}$, $\swap{2}_{2i}$, $\swap{10}_{2i}$, $\swap{(20-i)}_{2i}$ (co-terminal swap rate), $\mathcal{I}[\,\swap{2}_{2i}\le\swap{10}_{2i}\,]$, and $\mathcal{I}[\,0\le F_{2i}\le 0.03\,]$); and (ii) the four volatility parameters, $a$, $b$, $c$, and $d$ at $t_i$. With these terms, we construct the four basis sets in increasing order of $M$:\footnote{All basis sets include 1 for the intercept.}
\begin{itemize}
	\item $M=11$: linear terms of both groups.
	\item $M=28$: linear and quadratic terms of the interest rate group.
	\item $M=32$: linear and quadratic terms of the interest rate group and linear volatility terms.
	\item $M=66$: linear and quadratic terms of all variables in both groups.
\end{itemize}

In Table~\ref{tab:cmsra}, we present the pricing result. Similar to the previous examples, the LOOLSM prices are very close to the LSM-2 prices, indicating that the LOOLSM method removes look-ahead bias efficiently. The look-ahead bias, measured as the difference between the LSM and LOOLSM prices, proportionally increases as the number of basis $M$ increases. Similar to the previous cases, the LOOLSM price increases as more basis functions are used and suboptimal bias is reduced. Because the LOOLSM price is a low estimation, it is safe to use higher-order basis functions under the LOOLSM algorithm.

\begin{table}[!htb]
	\caption{\label{tab:cmsra} Results for the cancellable exotic swap (Case 4). The prices are for the notional value of 100. The prices are the average of the $n_\textsc{mc}=200$ simulation results with $N=40,000$ paths. The standard deviations are constant, around 0.12. The price of the underlying swap is $-4.22 \pm 0.09$. All values are rounded to two decimal places.}
	\centering
	\vspace{1ex}
	\begin{tabular}{|c||c|c|c||c|}
		\hline
		& \hspace{1em} LSM \hspace{1em} & \hspace{1ex} LSM-2 \hspace{1ex} & LOOLSM & LSM $-$ LOOLSM \\ \hline
		$M=11$ & 9.06 & 9.05 & 9.05 & 0.01 $\pm$ 0.003 \\
		$M=28$ & 9.42 & 9.40 & 9.40 & 0.02 $\pm$ 0.006 \\
		$M=32$ & 9.44 & 9.41 & 9.41 & 0.02 $\pm$ 0.005 \\
		$M=66$ & 9.49 & 9.45 & 9.44 & 0.05 $\pm$ 0.009 \\
		\hline
	\end{tabular}
\end{table}

\begin{table}[!htb]
	\caption{\label{tab:cmsra_time} Average computation time in seconds for pricing the cancellable exotic swap (Case 4) with $N=40,000$ paths ($2N$ paths for LSM-2) and $M=66$ basis functions. For smaller basis sets ($M=11$, 28, and 32), the ``Regression and Pricing'' time is smaller in proportion to $M$, while the ``Path Generation'' time is unchanged.
	} \centering \vspace{1ex}
	\begin{tabular}{|c||R{1.5cm}|R{1.5cm}|R{1.5cm}|} \hline
		& \multicolumn{1}{c|}{LSM} & \multicolumn{1}{c|}{LSM-2} & \multicolumn{1}{c|}{LOOLSM} \\ \hline
		Path Generation & 132.91 & 265.82 & 132.91 \\
		Regression and Pricing & 1.31 & 2.37 & 1.62 \\ \hline
		Total & 134.22 & 268.19 & 134.53 \\
		\hline
	\end{tabular}
\end{table}

In Table~\ref{tab:cmsra_time}, we compare the computation time of the three methods. We separately measure the time to generate paths and perform the regression and valuation. Note that path generation takes the majority of the total pricing time due to the complexity of the stochastic LIBOR market implementation. Since LSM-2 requires another set of simulation paths, generating paths takes twice as long. Regarding the time for regression and valuation, the increment from LSM to LOOLSM is marginal, as in \S~\ref{ssec:loolsm}, but LSM-2 takes longer than LOOLSM because we must evaluate the regression on the two simulation sets. Overall, the computational gain of the LOOLSM method is significant compared to the LSM-2 method, while they achieve the same goal of removing look-ahead bias. 

\section{Extension of LOOLSM to the other regression estimators} \label{sec:ext} \noindent
Many studies have aimed to improve the LSM method using advanced regression methods, such as ridge regression~\citep{tompaidis2014altols}, least absolute shrinkage and selection operator (LASSO)~\citep{tompaidis2014altols,chen2019var}, weighted least squares regression~\citep{fabozzi2017hetero,ibanez2018opt}, and non-parametric kernel regression~\citep{belomestny2011pricing,ludkovski2018kriging}. The LOOLSM method can be flexibly extended to these alternatives to the LSM method because they are essentially linear projections via the hat matrix. As long as $\lsv{h}{i}$ is available, the corrections from the original regression are obtained from \eqref{eq:2.3} in the same manner, and the two-pass approach becomes unnecessary.
Below, we present their hat matrices with discussions.

Ridge regression and LASSO are linear regressions with $L^2$ and $L^1$-regularization, respectively ~\citep[\S~3.4]{hastie2009ESL}. These methods outperform the LSM method in small simulation paths~\citep{tompaidis2014altols} and provide stable estimates of the value-at-risk~\citep{chen2019var}. The hat matrix of ridge regression is 
$$\lsv{H}{i} = \mat{X} (\mat{X}^\top \mat{X} + \lambda \mat{I}_M)^{-1} \mat{X}^\top,
$$
where $\lambda$ is the regularization strength. From its diagonal, we can see the implication of regularization for look-ahead bias. The effective degree of freedom, defined as the sum of the leverages (see \citet[(3.50)]{hastie2009ESL}), is less than that of the OLS regression in \eqref{eq:h_bound}:
$$ 
\sum_{n=1}^{N} \lsvel{h}{i}{n} = \sum_{j=1}^{M} \frac{d_j^2}{d_j^2 + \lambda} < M,
$$
where $d_j$ is the $j$-th singular value of $\lsv{X}{i}$. Therefore, we expect the look-ahead bias to decrease as $\lambda$ increases, following Section~\ref{sec:labias}. However, regularization alone cannot remove look-ahead bias completely, and we still need an additional method, such as the LOOLSM method. The hat matrix of LASSO is not analytically available, not to mention that the method is not exactly a linear projection. Given the selected regressors from shrinkage, however, it is a linear projection. Therefore, an approximation of the hat matrix can be obtained accordingly. The effective degree of freedom is equal to the number of the selected regressors under the approximation, which also indicates that LASSO has an effect of reducing look-ahead bias to some extent.

In the weighted linear regression, the hat matrix is
$$ \lsv{H}{i} = \mat{X} (\mat{X}^\top \mat{W} \mat{X})^{-1} \mat{X}^\top \mat{W},
$$
where $\mat{W}$ is an $N$-by-$N$ diagonal weight matrix. \citet{fabozzi2017hetero} adopt this approach to deal with heteroscedasticity and \citet{ibanez2018opt} to give higher weights to the paths near the exercise boundary. 

Despite heavy computation, the non-parametric kernel regression is an alternative to the OLS regression~\citep{belomestny2011pricing,ludkovski2018kriging}. For kernel function $\mathcal{K}(x,y)$, the hat matrix is the normalized kernel value between sample points. The $(n,n')$ element of $\lsv{H}{i}$ is 
$$\lsvel{H}{i}{n,n'} = \frac{\mathcal{K}(\lsvel{X}{i+1}{n},\lsvel{X}{i+1}{n'})}{\sum_{k=1}^{N} \mathcal{K}(\lsvel{X}{i+1}{n},\lsvel{X}{i+1}{k})}.
$$
The adjusted prediction value from \eqref{eq:2.3} is simply the self-excluded kernel estimate:
$$ \lsvel{C}{i}{n,\LOOLSM} = \frac{\sum_{k\neq n} \mathcal{K}(\lsvel{X}{i+1}{n},\lsvel{X}{i+1}{k})\, \lsvel{Y}{i+1}{k}}{\sum_{k\neq n} \mathcal{K}(\lsvel{X}{i}{n},\lsvel{X}{i}{k})}.
$$
In the kernel regression, the LOOLSM algorithm not only saves out-of-sample path generation but also saves costly kernel evaluations. \vspace{1em}

\section{Conclusion} \label{sec:con} \noindent
This study shows that it is possible to eliminate undesirable look-ahead bias in the LSM method \citep{longstaff2001val} using LOOCV without extra simulations. By measuring look-ahead bias with the LOOLSM method, we also find that the bias size is asymptotically proportional to the regressors-to-paths ratio. With numerical examples, we demonstrate that the LOOLSM method effectively prevents the possible overvaluation of multiasset Bermudan options without extra computation.

\section*{Conflict of Interest Statement} \noindent
The authors declare no conflict of interest.

\section*{Data Availability Statement} \noindent
Data sharing is not applicable to this article as it describes entirely theoretical research. No empirical data sets were analyzed or generated during the current study.

\appendix
\section{Proof of Theorem~\ref{theorem:3.1}} \noindent
We first introduce two technical Lemmas~\ref{lemma:A.1} and \ref{lemma:A.2}, then prove the main Theorem~\ref{theorem:3.1} in Section~\ref{sec:labias}. For ease of notation, we omit the exercise time superscripts $[i]$ from $\lsvel{Z}{i}{n}$, $\lsvel{C}{i}{n}$, $\lsvel{h}{i}{n}$, and $\lsvel{B}{i}{n}$ when the context is clear. However, we preserve the superscript in $\lsvel{Y}{i}{n}$ and $\lsvel{Y}{i+1}{n}$ to avoid ambiguity.

\label{apdx:deriveasympt}
\begin{lemma} \label{lemma:A.1}
	Let $\mathcal{I}[\,\cdot\,]$ be the indicator function. Then,
	$$\lsvel{B}{i}{n} \leq \mathcal{I}\left[\,|\lsvel{C}{}{n,\LSM} - \lsvel{Z}{}{n}| \le \lsvel{h}{}{n}\, |\lsvel{Y}{i+1}{n} - \lsvel{Z}{}{n}|\,\right]\; |\lsvel{Y}{i+1}{n} - \lsvel{Z}{}{n}|+\lsvel{B}{i+1}{n}. $$
\end{lemma}

\begin{proof}[Proof of Lemma~\ref{lemma:A.1}.] ~\\
We first formulate the conditions under which look-ahead bias changes the exercise decision. From (\ref{eq:modified}),
\begin{eqnarray*}
	\lsvel{B}{i}{n} &=& \lsvel{Y}{i}{n} - \lsvel{Y}{i}{n, \LOOLSM}\\
	&=& ( \mathcal{I}[\lsvel{Z}{}{n} \le \lsvel{C}{}{n,\LSM}] - \mathcal{I}[\lsvel{Z}{}{n} \le \lsvel{C}{}{n,\LOOLSM}]) (\lsvel{Y}{i+1}{n} - \lsvel{Z}{}{n}) + \mathcal{I}[\lsvel{Z}{}{n} \le \lsvel{C}{}{n,\LOOLSM}]\,\lsvel{B}{i+1}{n} \\
	&\le& (\mathcal{I}[D^+_n]-\mathcal{I}[D^-_n])(\lsvel{Y}{i+1}{n} - \lsvel{Z}{}{n}) + \lsvel{B}{i+1}{n}.
\end{eqnarray*}
where $D^+_n \iff \{\lsvel{C}{i}{n,\LOOLSM} < \lsvel{Z}{}{n} \le \lsvel{C}{i}{n,\LSM}\}$ and $D^-_n \iff \{\lsvel{C}{i}{n,\LSM} < \lsvel{Z}{}{n} \le \lsvel{C}{i}{n,\LOOLSM} \}$. Here, $D^+_n$ ($D^-_n$) is the condition in which the LSM algorithm incorrectly continues (exercises) due to look-ahead bias, but the LOOLSM algorithm exercises (continues), and the term $(\lsvel{Y}{i+1}{n} - \lsvel{Z}{}{n})$, is the price change caused by the inverted exercise decision. From (\ref{eq:2.3}), we obtain the following equivalence:
\begin{eqnarray*}
	D^+_n &\iff& 0 \le \lsvel{C}{}{n,\LSM} - \lsvel{Z}{}{n} < \lsvel{C}{}{n,\LSM} - \lsvel{C}{}{n,\LOOLSM} \\
	&\iff& 0 \le \lsvel{C}{}{n,\LSM} - \lsvel{Z}{}{n} < \frac{h_n}{1-h_n} (\lsvel{Y}{i+1}{n} - \lsvel{C}{}{n,\LSM}) \\
	&\iff& 0 \le \lsvel{C}{}{n,\LSM} - \lsvel{Z}{}{n} < h_n (\lsvel{Y}{i+1}{n} - \lsvel{Z}{}{n});
\end{eqnarray*}
similarly, 
$$D^-_n \iff h_n (\lsvel{Y}{i+1}{n} - \lsvel{Z}{}{n}) \le \lsvel{C}{}{n,\LSM} - \lsvel{Z}{}{n} < 0.$$ Since $D^+_n$ and $D^-_n$ are mutually exclusive events, we get
\begin{align*}
\lsvel{B}{i}{n} & \le \mathcal{I}[D^+_n \cup D^-_n] \cdot |\lsvel{Y}{i+1}{n} - \lsvel{Z}{}{n}| + \lsvel{B}{i+1}{n} \\ & \leq \mathcal{I}\left[\,|\lsvel{C}{}{n,\LSM} - \lsvel{Z}{}{n}| \le h_n |\lsvel{Y}{i+1}{n} - \lsvel{Z}{}{n}|\,\right]\cdot |\lsvel{Y}{i+1}{n} - \lsvel{Z}{}{n}| + \lsvel{B}{i+1}{n}. \quad\square
\end{align*}
\end{proof}

\begin{lemma} \label{lemma:A.2} 
	The following hold under Assumptions~\ref{assumption:1} and \ref{assumption:2}.
	\begin{enumerate}[(i)] \normalfont
		\item $\lsvel{Y}{i+1}{n} - \lsvel{Z}{}{n} \sim O_p(1)$ and $\ExpMC\left[ (\lsvel{Y}{i+1}{n} - \lsvel{Z}{}{n})^2 \right]$ are finite.
		\item $\lsvel{h}{}{n} \sim O_p(M/N)$.
		\item $|\lsvel{C}{}{n,\LSM} - \lsvel{Z}{}{n}|^{-1} \sim O_p(1)$.
	\end{enumerate}
\end{lemma}

\begin{proof}[Proof of Lemma~\ref{lemma:A.2}.]
\begin{enumerate}[(i)]
	\item Since $\lsvel{Y}{i+1}{n} = \lsvel{Z}{\tau}{n}$ for some $i+1 \le \tau\le I $, we obtain
	$$
	\left(\lsvel{Y}{i+1}{n} - \lsvel{Z}{i}{n}\right)^2 \le \left(\sum_{j=i}^I |\lsvel{Z}{j}{n} |\right)^2 \le (I-i+1) \; \sum_{j=i}^I \left(\lsvel{Z}{j}{n} \right)^2
	$$
	for any $N$. Then, the statements follow from Assumption~\ref{assumption:1} that $\lsf{Z}{i}(s)$ is in $L^2$.
	\item For any given $\varepsilon > 0$, 
	$$
	\ProbMC\left[h_n > \frac{1}{\varepsilon} \frac{M}{N}\right] \le \frac{\varepsilon N}{M}\; \ExpMC\left[h_n\right] < \varepsilon 
	$$
	using Markov's inequality and $\ExpMC[h_n]=M/N$.
	\item It is sufficient to prove that for any given $\varepsilon>0$, there exists $c>0$ such that $\ProbMC\left[\,|\lsvel{C}{}{n,\LSM} - \lsvel{Z}{}{n}| < c\,\right] < \varepsilon$ for any sufficiently large $N$. From Assumption~\ref{assumption:2}, we can choose $c$ so that
	$$ \ProbMC\left[\,|C_{n,M} - \lsvel{Z}{}{n}| < 2c \;\right] < \frac{\varepsilon}{2}\;,
	$$
	From \citet[Lemma 3.2]{clement2002anal}, which is also based on Assumption~\ref{assumption:2}, $\lsf{\hat{C}}{i}_\LSM(s)$ almost surely converges to $C_{M}(s)$ for a large $N$:
	$$ \ProbMC\left[\,|\lsvel{C}{}{n,\LSM} - C_{n, M}| \ge c \;\right] < \frac{\varepsilon}{2} .
	$$
	Here, the choice of $c$ does not depend on $M$ because Assumption~\ref{assumption:2} holds uniformly on $M$. Then,
	\begin{align*}
	\ProbMC\left[\,|\lsvel{C}{}{n,\LSM} - \lsvel{Z}{}{n}| < c\,\right] 
	&\le \ProbMC\left[\, \{|\lsvel{C}{}{n,\LSM} - \lsvel{Z}{}{n}| < c\} \cap \{|\lsvel{C}{}{n,\LSM} - C_{n, M}| < c\}\,\right]
	+ \ProbMC\left[\,|\lsvel{C}{}{n,\LSM} - C_{n, M}| \ge c\,\right]\\ 
	&< \ProbMC\left[\,|C_{n, M} - \lsvel{Z}{}{n}| < 2c\;\right] + \ProbMC\left[\,|\lsvel{C}{}{n,\LSM} - C_{n, M}| \ge c\,\right]\\
	&= \frac{\varepsilon}{2} + \frac{\varepsilon}{2} = \varepsilon. \quad\square
	\end{align*}	
\end{enumerate}
\end{proof}

\thmepsilon*
\proof{Proof of Theorem~\ref{theorem:3.1}.}
\begin{enumerate}[(i)]
	\item We prove this Theorem inductively. First, $\lsvel{B}{I}{n} = 0$, and we assume that $\lsvel{B}{i+1}{n} \sim O_p(M/N)$ for $i < I$. We define $E_n, F_n$ and $G_n$ as
	\begin{align*}
	E_n &\iff \{h_n > kM/N\} \; \cup \; \{|\lsvel{C}{}{n,\LSM} - \lsvel{Z}{}{n}| < c\}, \\
	F_n &\iff \{|\lsvel{Y}{i+1}{n} - \lsvel{Z}{}{n}| > l\}, \\
	G_n &\iff \{\lsvel{B}{i+1}{n} > s M/N\}.
	\end{align*}
	For $\varepsilon > 0$, we choose $c$, $k$, $l$, and $s$ such that $\ProbMC[E_n \cup F_n \cup G_n] < \varepsilon$ by Lemma~\ref{lemma:A.2} and the induction assumption. If $\mathcal{I}[E_n \cup F_n \cup G_n] = 0$, then
	\begin{align*}
	\lsvel{B}{i}{n} &\le \mathcal{I}\left[\,|\lsvel{C}{}{n,\LSM} - \lsvel{Z}{}{n}| \le h_n |\lsvel{Y}{i+1}{n} - \lsvel{Z}{}{n}|\,\right] \, |\lsvel{Y}{i+1}{n} - \lsvel{Z}{}{n}| + \lsvel{B}{i+1}{n} \\
	&\le \mathcal{I}\left[c \le \frac{kM}{N} |\lsvel{Y}{i+1}{n} - \lsvel{Z}{}{n}|\,\right]\, |\lsvel{Y}{i+1}{n} - \lsvel{Z}{}{n}| + \lsvel{B}{i+1}{n} \le \left(\frac{kl^2}c + s\right)\frac{M}{N}
	\end{align*}
	by Lemma~\ref{lemma:A.1} and Markov's inequality. Finally, 
	$$
	\ProbMC\left[\lsvel{B}{}{n} > \left(\frac{kl^2}c + s\right)\frac{M}{N} \right] \le \ExpMC\left[\,
	\mathcal{I}[E_n \cup F_n \cup G_n]\, \right] < \varepsilon.
	$$
	Therefore, $\lsvel{B}{i}{n} \sim O_p(M/N)$. 
	
	\item 
	By Lemma~\ref{lemma:A.2}(i), $\ExpMC[(Y_n - Z_n)^2] = L$ for some $L>0$. Let $\eta = \frac{\varepsilon^2}{L(I-1)^2}$ for any given $\varepsilon>0$. By Lemma~\ref{lemma:A.2}(ii) and (iii), we can choose $c$ and $k$ such that 
	$$\ProbMC\left[\,h_n > \frac{k M}{N}\,\right] < \frac{\eta}2 \qtext{and} \ProbMC[\,|\lsvel{C}{}{n,\LSM} - Z_n|<c\,] < \frac{\eta}2.$$
	Then, 
	$$\ProbMC[E_n]\,\ExpMC[(\lsvel{Y}{i+1}{n} - \lsvel{Z}{}{n})^2] < \frac{\varepsilon^2}{(I-1)^2}
	$$ for any sufficiently large $N$. Then,
	\begin{align*}
	\ExpMC[\,\lsvel{B}{i}{n} & - \lsvel{B}{i+1}{n}] \le \ExpMC \left[
	\mathcal{I}[\,|\lsvel{C}{}{n,\LSM} - \lsvel{Z}{}{n}| \le h_n |\lsvel{Y}{i+1}{n} - \lsvel{Z}{}{n}|\,]\cdot |\lsvel{Y}{i+1}{n} - \lsvel{Z}{}{n}|\, \right] \tag{by Lemma~\ref{lemma:A.1}}\\
	&\le \ExpMC \left[ \mathcal{I}[ E_n ] \cdot |\lsvel{Y}{i+1}{n} - \lsvel{Z}{}{n}|\,\right] \\
	&\qquad + \ExpMC \left[ \mathcal{I}[E_n^c \cap (|\lsvel{C}{i}{n,\LSM} - \lsvel{Z}{}{n}| \le h_n |\lsvel{Y}{i+1}{n} - \lsvel{Z}{}{n}|) ] \cdot |\lsvel{Y}{i+1}{n} - \lsvel{Z}{}{n}|\, \right] \\
	&\le \ExpMC \left[ \mathcal{I}[ E_n ] \cdot |\lsvel{Y}{i+1}{n} - \lsvel{Z}{}{n}|\,\right] +
	\ExpMC \left[ \mathcal{I}\!\left[\,c \le \frac{kM}{N} |\lsvel{Y}{i+1}{n} - \lsvel{Z}{}{n}|\,\right]\cdot |\lsvel{Y}{i+1}{n} - \lsvel{Z}{}{n}| \,\right] \\
	&\le \ExpMC \left[ \mathcal{I}[ E_n ] \cdot |\lsvel{Y}{i+1}{n} - \lsvel{Z}{}{n}|\,\right] +
	\frac{kM}{cN} \ExpMC \left[ (\lsvel{Y}{i+1}{n} - \lsvel{Z}{}{n})^2 \right] \tag{by Markov's inequality}\\
	&\le \Big(\ProbMC\left[E_n\right]\, \ExpMC \left[ (\lsvel{Y}{i+1}{n} - \lsvel{Z}{}{n})^2 \right]\Big)^{1/2} +
	\frac{kM}{cN} \ExpMC \left[ (\lsvel{Y}{i+1}{n} - \lsvel{Z}{}{n})^2 \right] \tag{by the Cauchy--Schwarz inequality}\\
	&< \frac{\varepsilon}{I-1} + \frac{M}{N}r_{\varepsilon}^{[i]}. \tag{for some $r_{\varepsilon}^{[i]}$, by Lemma~\ref{lemma:A.2}(i)}
	\end{align*}
	The second inequality above is obtained from 
	$$\mathcal{I}[A] = \mathcal{I}[E_n \cap A] + \mathcal{I}[E_n^c \cap A] \le \mathcal{I}[E_n] + \mathcal{I}[E_n^c \cap A]$$
	for any event $A$, where $E_n^c$ is the complementary event of $E_n$.

	Finally, we can aggregate the step-wise bounds for the incremental bias to obtain an upper bound for the overall bias:
	$$
	\ExpMC[\hat{B}] = \ExpMC[\,\ExpN [\lsvel{B}{1}{n}]\,] = \ExpMC[\lsvel{B}{1}{n}] = \sum_{i=1}^{I-1} \ExpMC[\lsvel{B}{i}{n} - \lsvel{B}{i+1}{n}] < \varepsilon + \frac{M}{N}\sum_{i=1}^{I-1} r_{\varepsilon}^{[i]}.
	$$
	This completes the proof.
	\item For any given $\delta$ and $\varepsilon$, we can choose $\varepsilon_1$ and $\varepsilon_2$ such that $\ExpMC[\lsvel{B}{1}{n}] \le \varepsilon_1 + r_{\varepsilon_1}M/N < \delta \varepsilon$ for any $M/N<\varepsilon_2$. Then,
	$$
	\ProbMC[\hat{B}>\delta\,] \le \frac{1}{\delta} \ExpMC[ \hat{B}\,] = \frac{1}{\delta} \ExpMC [\lsvel{B}{1}{n}] < \varepsilon.
	$$
	Therefore, $\hat{B}$ converges to zero in probability. $\square$
\end{enumerate}
\endproof

\newpage
\singlespacing
\bibliography{LOOLSM_Z}

\end{document}